\documentclass[runningheads]{llncs}
\usepackage{graphicx}
\usepackage[dvipsnames]{xcolor}

\usepackage{float}
\usepackage{mathtools}
\usepackage{amsmath,amssymb}
\usepackage{stmaryrd}
\usepackage{multirow}
\usepackage{microtype}
\usepackage{caption,subcaption}

\usepackage{tikz}
\usepackage{pgfplots}
\usetikzlibrary{matrix,shapes}
\usepgfplotslibrary{groupplots}

\usepackage{mdframed}

\usepackage[misc]{ifsym}

\usepackage[algoruled, linesnumbered, noend]{algorithm2e}

\usepackage{xspace}

\usepackage[firstpage]{draftwatermark}

\newcommand{\Act}{Act}
\newcommand{\act}{\ensuremath{\alpha}}
\newcommand{\Obs}{Z}
\newcommand{\obs}{z}

\newcommand{\iversion}[1]{\ensuremath{\left[#1\right]}}
\newcommand{\naturals}{\mathbb{N}}

\newcommand{\unitinterval}{[0,1]}
\newcommand{\supp}{\mathrm{supp}}

\newcommand{\sinit}{s_{0}}
\newcommand{\mpm}{\mathcal{P}}
\newcommand{\mdpT}{(S,\sinit,Act,\mpm)}

\newcommand{\pomdp}{\mathcal{M}}
\newcommand{\pomdpT}{(M,Z,O)}

\newcommand{\ninit}{n_{0}}
\newcommand{\fscT}{(N,\ninit,\gamma,\delta)}

\newcommand{\fsc}{F}
\newcommand{\imc}{\pomdp^\fsc}
\newcommand{\fimc}{\pomdp^{\ffsck}}
\newcommand{\fmdp}{\textsf{MDP}({\ffsck})}

\newcommand{\ffsc}{\mathcal{F}^\pomdp}
\newcommand{\ffsck}{{\ffsc_k}}
\newcommand{\ffscmu}{{\ffsc_{\mu}}}

\newcommand{\belief}{\mathcal{B}}
\newcommand{\beliefs}{\belief_\pomdp}
\newcommand{\bsucc}[3]{\llbracket #1 {\mid} #2,#3\rrbracket}
\newcommand{\binit}{b_{0}}
\newcommand{\bpm}{\mpm^\belief}
\newcommand{\bmdp}{\pomdp^{\belief}}
\newcommand{\actions}{\ensuremath{Act}}
\newcommand{\bmdpT}{(\beliefs, \binit, \actions, \bpm)}
\newcommand{\btarget}{T^\belief}

\newcommand{\bsigma}{{\sigma_{\belief}}}

\newcommand{\F}[1]{\Diamond #1}
\newcommand{\prob}[2][]{\mathbb{P}^{#1}\left[#2\right]}
\newcommand{\probmax}[2][]{\mathbb{P}^{#1}_{\max}\left[#2\right]}
\newcommand{\probmin}[2][]{\mathbb{P}^{#1}_{\min}\left[#2\right]}

\newcommand{\reach}[1]{\mathbb{P}^{\pomdp^{#1}}\!\left[\Diamond{T}\right]}
\newcommand{\statereach}[2]{\mathbb{P}^{\pomdp^{#1}}\!\left[#2 \models \Diamond{T}\right]}

\newcommand{\bsched}{\sigma^{\bmdp}}

\newcommand{\clipmdp}{\overline{\bmdp}}
\newcommand{\approximator}{\overline{F}}
\newcommand{\btop}{b_{\top}}
\newcommand{\bbot}{b_{\bot}}

\newcommand{\inductive}{\mathcal{I}}
\newcommand{\design}{\mathcal{D}}
\newcommand{\designmu}{\design_{\mu}}

\newcommand{\fscind}{\ensuremath{F_{\inductive}}\xspace}
\newcommand{\fscbel}{\ensuremath{F_{\belief}}\xspace}
\newcommand{\lb}{\underline{V}}

\newcommand{\exploredbel}{\ensuremath{\mathcal{E}}}
\newcommand{\frontier}{\ensuremath{\mathcal{U}}}

\newcommand{\tool}[1]{{\textsc{#1}}}
\newcommand{\storm}{\tool{Storm}\xspace}
\newcommand{\paynt}{\tool{Paynt}\xspace}

\newcommand{\integration}{\tool{Saynt}\xspace}

\newcommand{\size}{size}
\newcommand{\nodes}[1]{N_{#1}}

\makeatletter
\RequirePackage[bookmarks,unicode,colorlinks=true,breaklinks=true]{hyperref}%
   \def\@citecolor{blue}%
   \def\@urlcolor{blue}%
   \def\@linkcolor{blue}%

\def\orcidID#1{\smash{\href{http://orcid.org/#1}{\protect\raisebox{-1.25pt}{\protect\includegraphics{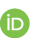}}}}}
\makeatother

\renewcommand{\paragraph}[1]{\smallskip\noindent\emph{#1}}
\renewcommand{\subsubsection}[1]{\medskip\noindent\textbf{#1}}

\begin{document}
\title{Search and Explore:\\Symbiotic Policy Synthesis in POMDPs\thanks{This work has been supported by the Czech Science Foundation grant \mbox{GA23-06963S} (VESCAA), the ERC AdG Grant \mbox{787914} (FRAPPANT) and the DFG RTG 2236/2 (UnRAVeL).}
}
%
%
\author{Roman Andriushchenko\inst{1}\orcidID{0000-0002-1286-934X} \and Alexander Bork \inst{2}\orcidID{0000-0002-7026-228X} \and
Milan \v{C}e\v{s}ka {(\Letter)} \inst{1}\orcidID{0000-0002-0300-9727} \and\\
Sebastian Junges\inst{3}\orcidID{0000-0003-0978-8466} \and 
Joost-Pieter Katoen\inst{2}\orcidID{0000-0002-6143-1926} \and
Filip Mac\'{a}k\inst{1}\orcidID{0009-0004-4277-2751}}

\authorrunning{R. Andriushchenko et al.}

\institute{
Brno University of Technology, Brno, Czech Republic \\
\email{ceskam@fit.vutbr.cz}
\and
RWTH Aachen University, Aachen, Germany
\and
Radboud University, Nijmegen, The Netherlands
}

\SetWatermarkAngle{0}
\SetWatermarkText{} 
\SetWatermarkText{\raisebox{11.5cm}{%
  \hspace{0.1cm}%
  \href{https://doi.org/10.5281/zenodo.7874513}{\includegraphics{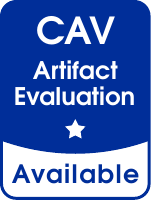}}%
  \hspace{9cm}%
  \includegraphics{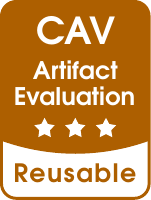}%
}}



\maketitle
\begin{abstract}
This paper marries two state-of-the-art controller synthesis methods for partially observable Markov decision processes (POMDPs), a prominent model in sequential decision making under uncertainty.
A central issue is to find a POMDP controller---that solely decides based on the observations seen so far---to achieve a total expected reward objective.
As finding optimal controllers is undecidable, we concentrate on synthesising good finite-state controllers (FSCs).
We do so by tightly integrating two modern, orthogonal methods for POMDP controller~synthesis: a belief-based and an inductive approach. 
The former method obtains an FSC from a finite fragment of the so-called belief MDP, an MDP that keeps track of the probabilities of equally observable POMDP states.
The latter is an inductive search technique over a set of FSCs, e.g., controllers with a fixed memory size.
The key result of this paper is a symbiotic anytime algorithm that tightly integrates both approaches such that each profits from the controllers constructed by the other.
Experimental results indicate a substantial improvement in the value of the controllers while significantly reducing the synthesis time and memory~footprint.

\end{abstract}

\section{Introduction} 

A formidable synthesis challenge is to find a decision-making policy that satisfies temporal constraints even in the presence of stochastic noise. 
\emph{Markov decision processes (MDPs)}~\cite{puterman1994} are a prominent model to reason about such policies under stochastic uncertainty. 
The underlying decision problems are efficiently solvable and probabilistic model checkers such as PRISM~\cite{DBLP:conf/cav/KwiatkowskaNP11} and \storm~\cite{STORM} are well-equipped to synthesise policies that provably (and optimally) satisfy a given specification. 
However, a major shortcoming of MDPs is the assumption that the policy can depend on the precise state of a system. 
This assumption is unrealistic whenever the state of the system is only observable via sensors. 
\emph{Partially observable MDPs (POMDPs)} overcome this shortcoming, but policy synthesis for POMDPs and specifications such as \emph{the probability to reach the exit is larger than 50$\%$} requires solving undecidable problems~\cite{madani2005undecidability}. 
Nevertheless, in recent years, a variety of approaches have been successfully applied to a variety of challenging benchmarks, but the approaches also fail somewhat spectacularly on seemingly tiny problem instances. 
From a user perspective, it is hard to pick the right approach without detailed knowledge of the underlying methods. This paper sets out to develop a framework in which conceptually orthogonal approaches symbiotically alleviate each other's weaknesses and find policies that maximise, e.g., the expected reward before a target is reached. We show empirically that the combined approach can find compact policies achieving a significantly higher reward than the policies that either individual approach constructs.

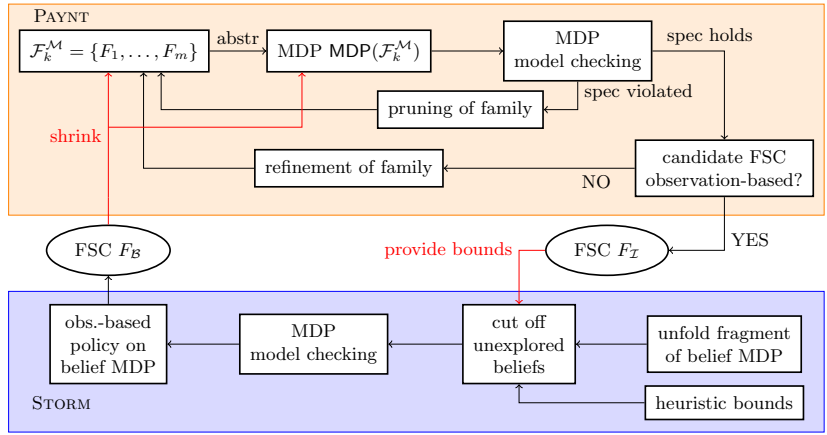
\begin{figure}[tb]
\centering
\scalebox{0.78}{
\begin{tikzpicture}[font=\footnotesize]
\draw[draw=orange, fill=orange!15] (-1.7,0.8) rectangle (12.2,-2.8);
\node[] at (-0.8,0.6) {\paynt};
\draw[draw=blue, fill=blue!15] (-1.7,-4.1) rectangle (12.2,-6.5);
\node[] at (-0.8,-6) {\tool{Storm}};
\node[draw, fill=white, thick, rectangle, inner sep=5, align=center] (fam) at (0.1,0) {$\ffsck = \{F_1,\dots,F_m\}$};
\node[draw, fill=white, thick, rectangle, inner sep=5, align=center] (abstmdp) at (4.1,0) {MDP $\fmdp$};
\node[draw, fill=white, thick, rectangle, inner sep=5, align=center] (payntmc) at (8,0) {MDP\\model checking};
\node[draw,fill=white, thick, rectangle, inner sep=5, align=center] (canfsc) at (10.5,-2) {candidate FSC\\observation-based?};
\node[draw,fill=white, thick, rectangle, inner sep=5, align=center] (famRefinement) at (4.1,-2) {refinement of family};
\node[draw,fill=white, thick, rectangle, inner sep=5, align=center] (bounds) at (6,-1) {pruning of family};
\node[inner sep=0] (payntentry) at (0,-2.5) {};
\node[draw, fill=white,thick, ellipse, inner sep=5, align=center] (payntfsc) at (8.5,-3.4) {FSC $\fscind$};
\node[draw, fill=white, thick, rectangle, inner sep=5, align=center] (beliefUnfold) at (10.5,-5) {unfold fragment\\of belief MDP};
\node[draw, fill=white, thick, rectangle, inner sep=5, align=center] (beliefCutoff) at (7,-5) {cut off\\unexplored\\beliefs};
\node[draw, fill=white, thick, rectangle, inner sep=5, align=center] (stormmc) at (3.5,-5) {MDP\\ model checking};
\node[draw, fill=white, thick, rectangle, inner sep=5, align=center] (stormpol) at (0,-5) {obs.-based\\policy on\\belief MDP};
\node[draw, fill=white, thick, rectangle, inner sep=5, align=center] (heuristic) at (10.5,-6) {heuristic bounds};
\node[draw, fill=white, thick, ellipse, inner sep=5, align=center] (stormfsc) at (0,-3.4) {FSC $\fscbel$};
\draw[->] (fam) -- node[above]{abstr} (abstmdp);
\draw[->] (abstmdp) -- (payntmc);
\draw[->] (payntmc) -| node[above, pos=0.4]{spec holds} (canfsc);
\draw[->] (payntmc) |- node[right, align=center,pos =0.2]{spec violated} (bounds);
\draw[->] (bounds) -| ([shift={(0.8,0)}]fam.south);
\draw[->] (canfsc) |- node[right, pos=0.4] {YES} (payntfsc);
\draw[->] (canfsc) -- node[below, pos=0.2] {NO} (famRefinement);
\draw[->] (famRefinement) -| ([shift={(0.5,0)}]fam.south);
\draw[->, color=red] (payntfsc) -| node[left] {\color{red}provide bounds} (beliefCutoff);
\draw[->] (beliefUnfold) -- (beliefCutoff);
\draw[->] (beliefCutoff) -- (stormmc);
\draw[->] (stormmc) -- (stormpol);
\draw[->] (stormpol) -- (stormfsc);
\draw[->] (heuristic) -| (beliefCutoff);
\draw[-, color=red] (stormfsc) -- (payntentry.center);
\draw[->, color=red] (payntentry.center) -- node[left] {\color{red}shrink} ([shift={(-0.1,0)}]fam.south);
\draw[->, color=red] ([shift={(0,1.2)}]payntentry.center) -| ([shift={(-0.8,0)}]abstmdp.south);
\end{tikzpicture}
}
\vspace{-0.5em}
\caption{Schematic depiction of the symbiotic approach}
\label{fig:saynt}
\vspace{-1em}
\end{figure}
\paragraph{Belief exploration.}
Several approaches for solving POMDPs use the notion of \emph{beliefs}~\cite{smallwood1973optimal}. 
The key idea is that each sequence of observations and actions induces a belief---a distribution over POMDP states that reflects the probability to be in a state conditioned on the observations.
POMDP policies can decide optimally solely based on the belief.
The evolution of beliefs can be captured by a fully observable, yet possibly infinite \emph{belief MDP}. 
A practical approach (see the lower part of Fig.~\ref{fig:saynt}) is to unfold a finite fragment of this belief MDP and make its frontier absorbing.
This finite fragment can be analysed with off-the-shelf MDP model checkers.
Its accuracy can be improved by using an arbitrary but fixed cut-off policy from the frontier onwards. 
Crucially, the probability to reach the target under such a policy can be efficiently pre-computed for all beliefs. 
This paper considers the belief exploration method from~\cite{bork2022under} realised in \storm~\cite{STORM}.

\paragraph{Policy search.}
An orthogonal approach searches a (finite) space of policies~\cite{DBLP:conf/uai/Hansen98,DBLP:conf/uai/MeuleauKKC99} and evaluates these policies by verifying the induced Markov chain. 
To ensure scalability, sets of policies must be efficiently analysed.
However, policy spaces explode whenever they require memory. The open challenge is to adequately define the space of policies to search in. 
In this paper, we consider the policy-search method from~\cite{andriushchenko2022inductive} as implemented in \paynt~\cite{andriushchenko2021paynt} that explores spaces of finite-state controllers (FSCs), represented as deterministic Mealy machines~\cite{DBLP:conf/aaai/AmatoBZ10}, using a combination of abstraction-refinement, counterexamples (to prune sets of policies), and increasing a controller's memory, see the upper part of Fig.~\ref{fig:saynt}. 

\paragraph{Our symbiotic approach.}
In essence, our idea relies on the fact that a policy found via one approach can boost the other approach. 
The key observation is that such a policy is beneficial even when it is sub-optimal in terms of the objective at hand.
Fig.~\ref{fig:saynt} sketches the symbiotic approach.
The FSCs \fscind obtained by policy search are used to guide the partial belief MDP to the target.
Vice versa, the FSCs \fscbel obtained from belief exploration are used to shrinken the set of policies and to steer the abstraction. 
Our experimental evaluation, using a large set of POMDP benchmarks, reveals that (a)~belief exploration can yield better FSCs (sometimes also faster) using FSCs \fscind from \paynt---even if the latter FSCs are far from optimal, (b)~policy search can find much better FSCs when using FSCs from belief exploration, and (c)~the FSCs from the symbiotic approach are superior in value to the ones obtained by the standalone approaches.

\paragraph{Beyond exploration and policy search.}
In this work, we focus on two powerful orthogonal methods from the set of belief-based and search-based methods. Alternatives exist. Exploration can also be done using a fixed set of beliefs~\cite{norman2017verification}. Prominently, HSVI~\cite{horak2018hsvi} and SARSOP~\cite{kurniawati2008sarsop} are belief-based policy synthesis approaches typically used for discounted properties. They also support undiscounted properties, but represent policies with $\alpha$-vectors. Bounded policy synthesis~\cite{DBLP:conf/atal/WangCK18} uses a combination of belief-exploration and inductive synthesis over paths and addresses finite horizon reachability. 
$\alpha$-vector policies lead to more complex analysis downstream: the resulting policies must track the belief and do floating-point computations to select actions.
For policy search, prominent alternatives are to search for randomised controllers via gradient descent~\cite{DBLP:conf/vmcai/HeckSJMK22} or via convex optimization~\cite{amato2010optimizing,junges2018finite,DBLP:conf/aaai/Cubuktepe0JMST21}. Alternatively, FSCs can be extracted via deep reinforcement learning~\cite{DBLP:journals/jair/Carr0T21}. However, randomised policies limit predictability, which hampers testing and explainability. The area of programmatic reinforcement learning~\cite{DBLP:conf/icml/VermaMSKC18} combines inductive synthesis ideas with RL.
While our empirical evaluation is method-specific, the lessons carry over to integrating other methods.

\paragraph{Contributions.}
The key contribution of this paper is the symbiosis of belief exploration~\cite{bork2022under} and policy search~\cite{andriushchenko2022inductive}.
Though this seems natural, various technical obstacles had to be addressed, e.g., obtaining~$\fscbel$ from the finite fragment of the belief MDP and the policies for its frontier and developing an interplay between the exploration and search phases that minimises the overhead.
The benefits of the symbiotic algorithm are manifold, as we show by a thorough empirical evaluation.
It can solve POMDPs that cannot be tackled with either of the two approaches alone.
It outputs FSCs that are superior in value (with relative improvements of up to 40\%)
as well as FSCs that are more succinct (with reduction of a factor of up to two orders of magnitude) with only a small penalty in their values. Additionally, the integration reduces the memory footprint compared to belief exploration by a factor of 4.
In conclusion, the proposed symbiosis offers a powerful push-button, anytime synthesis algorithm producing, in the given time, superior and/or more succinct FSCs compared to the state-of-the-art~methods.

\section{Motivating Examples}
\label{sec:motivation}
We give a sample POMDP that is hard for the belief exploration, a POMDP that challenges the policy search approach, and indicate why a symbiotic  approach overcomes this.
A third sample POMDP is shown to be unsolvable by either approach alone but can be treated by the symbiotic one.

\begin{figure}[t]
\centering
\begin{subfigure}{.38\textwidth}
\centering
\includegraphics[width=0.8\linewidth]{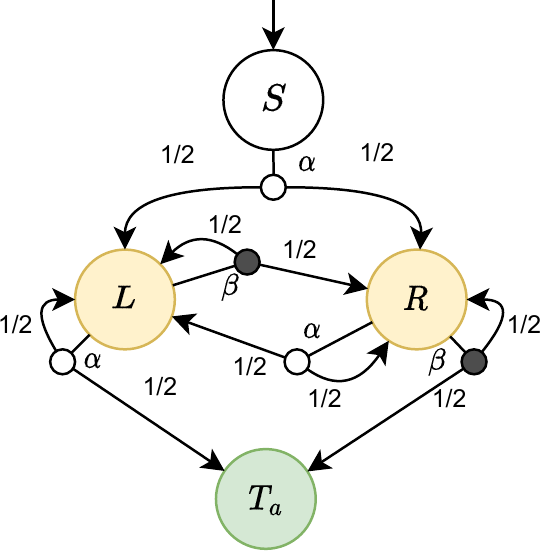}
\caption{}
\label{fig:storm-problem}
\end{subfigure}%
\begin{subfigure}{.6\textwidth}
\centering
\includegraphics[width=0.8\linewidth]{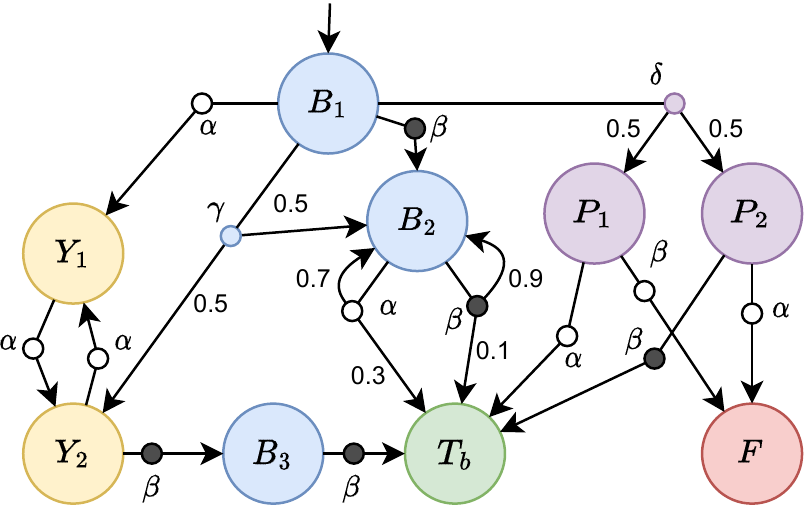}
\caption{}
\label{fig:paynt-problem}
\end{subfigure}%

\vspace{0.5em}
\begin{subfigure}{\textwidth}
\centering
\includegraphics[width=.85\linewidth]{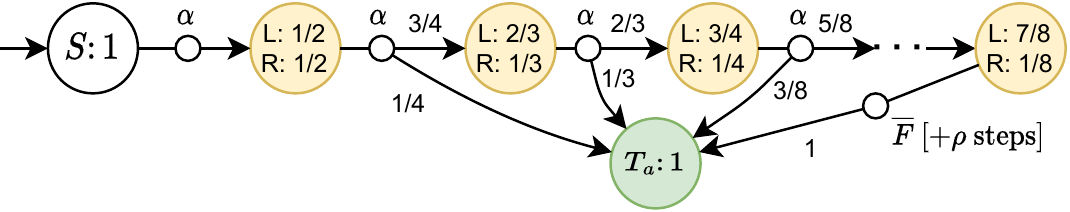}
\vspace{-2em}
\caption{}
\label{fig:randomized-cutoff}
\end{subfigure}%
\vspace{-0.5em}
\caption{(a) and (b) contain two POMDPs. Colours encode observations. Unlabelled transitions have probability 1. Omitted actions (e.g. $\gamma, \delta$ in state $B_2$) execute a self-loop. (c) Markov chain induced by the minimising policy $\bsigma$ in the finite abstraction $\overline{\bmdp_a}$ of the POMDP from Fig.~\ref{fig:storm-problem}. In the rightmost state, policy $\approximator$ is applied (cut-off), allowing to reach the target in $\rho$ steps.}
\label{fig:problems}
\end{figure}

\subsubsection{A challenging POMDP for belief-based exploration.}
Consider POMDP $\pomdp_a$ in Fig.~\ref{fig:storm-problem}. 
The objective is to minimise the expected number of steps to the target $T_a$. 
An optimal policy is to always take action $\alpha$ yielding 4 expected steps. 
An FSC realising this policy can be found by a policy search under 1s.

\paragraph{Belief MDPs.}
States in the \emph{belief MDP} $\bmdp_a$ are \emph{beliefs}, probability distributions over POMDP states with equal observations.
The initial belief is $\{S \mapsto 1\}$. 
By taking action $\alpha$, `yellow' is observed and the belief becomes $\{L \mapsto \frac{1}{2},\, R \mapsto \frac{1}{2}\}$. 
Closer inspection shows that the set of reachable beliefs is infinite rendering $\bmdp_a$ to be infinite.
Belief exploration constructs a finite fragment $\overline{\bmdp_a}$ by exploring $\bmdp_a$ up to some depth while \emph{cutting off} the frontier states. 
From cut-off states, a shortcut is taken directly to the target. 
These shortcuts are heuristic over-approximations of the true number of expected steps from the cut-off state to the target. 
The finite MDP $\overline{\bmdp_a}$ can be analysed using off-the-shelf tools yielding the minimising policy $\bsigma$ assigning to each belief state the optimal action.

\paragraph{Admissible heuristics.}
A simple way to over-approximate the minimal number of the expected number of steps to the target is to use an arbitrary controller $\approximator$ and use the expected number of steps under $\approximator$. 
The latter is cheap if $\approximator$ is compact, as detailed in Sec.~\ref{subsec:paynt-to-storm}. 
Fig.~\ref{fig:randomized-cutoff} shows a Markov chain induced by $\bsigma$ in $\overline{\bmdp_a}$, where the belief $\{ L \mapsto \frac{7}{8}, R \mapsto \frac{1}{8} \}$ is cut off using $\approximator$. 
The belief exploration in \storm~\cite{bork2022under} unfolds 1000 states of $\bmdp_a$ and finds controller $\approximator$ that uniformly randomises over all actions in the rightmost state. 
The resulting sub-optimal controller \fscbel reaches the target in ${\approx} 4.1$ steps.
Exploring only a few states suffices when replacing $\approximator$ by a (not necessarily optimal) FSC provided by a policy search.

\subsubsection{A challenging POMDP for policy search.}
Consider POMDP~$\pomdp_b$ in Fig.~\ref{fig:paynt-problem}. 
The objective is to minimise the expected number of steps to $T_b$.
Its 9-state belief MDP $\bmdp_b$ is trivial for the belief-based method.
Its optimal controller $\bsigma$ first picks action $\gamma$; on observing `yellow' it plays $\beta$ twice, otherwise it always picks $\alpha$. 
This is realised by an FSC with 3 memory states. 
The inductive policy search in \paynt~\cite{andriushchenko2022inductive} explores families of FSCs of increasing complexity, i.e., of increasing memory size. 
It finds the optimal FSC after consulting about 20~billion candidate policies. 
This requires 545 model-checking queries; the optimal one is found after 105 queries while the remaining queries prove that no better 3-state FSC exists.

\paragraph{Reference policies.}
The policy search is guided by a reference policy, in this case the fully observable MDP policy that picks (senseless) action $\delta$ in $B_1$ first. 
Using policy $\bsigma$---obtained by the belief method---instead, $\delta$ is never taken. 
As $\bsigma$ picks in each `blue' state a different action, mimicking this requires at least three memory states. 
Using $\bsigma$ reduces the total number of required model-checking queries by a factor of ten; the optimal 3-state FSC is found after 23 queries.

\subsubsection{The potential of symbiosis.}
To further exemplify the limitation of the two approaches and the potential of their symbiosis, we consider a synthetic POMDP, called Lanes+, combining a Lane model with larger variants of the POMDPs in Fig.~\ref{fig:problems}; see Tab.~\ref{tab:stats} on page~\pageref{tab:stats} for the model statistics and Appendix~C of~\cite{andriushchenko2023search} for the model description.
We consider minimisation of the expected number of steps and a 15-minute timeout. The belief-based approach by \storm yields the value 18870. 
The policy search method by \paynt finds an FSC with 2 memory states achieving the value 8223. 
This sub-optimal FSC significantly improves the belief MDP approximation and enables \storm to find an FSC with value 6471. 
The symbiotic synthesis loop finds the optimal FSC with value 4805.

\section{Preliminaries and Problem Statement}

A (discrete) \emph{distribution} over a countable set $A$ is a~function $\mu \colon A \rightarrow \unitinterval$ s.t.~$\sum_a \mu(a) = 1$.
The set $\supp(\mu) \coloneqq \left\{ a \in A \mid \mu(a) > 0\right\}$ is the \emph{support} of $\mu$.
The set $Distr(A)$ contains all distributions over $A$.
We use Iverson bracket notation, where $\iversion{x} = 1$ if the Boolean expression $x$ evaluates to true and $\iversion{x} = 0$ otherwise.

\begin{definition}[MDP]
A \emph{Markov decision process (MDP)} is a tuple $M = \mdpT$ with a countable set $S$ of states, an initial state $\sinit \in S$, a finite set $\Act$ of actions, and a partial transition function $\mpm \colon S \times \Act \nrightarrow Distr(S)$.
$\Act(s) \coloneqq \left\{ \act \in \Act \mid \mpm(s,\act) \neq \bot \right\}$ denotes the set of actions available in state $s \in S$.
An MDP with $|\Act(s)|=1$ for each $s \in S$ is a \emph{Markov chain (MC)}.

\end{definition}
Unless stated otherwise, we assume $\Act(s) = \Act$ for each $s \in S$ for conciseness.
We denote $\mpm(s,\act,s') \coloneqq \mpm(s,\act)(s')$.
A (finite) \emph{path} of an MDP $M$ is a sequence $\pi = s_0\act_0s_1\act_1\dots s_n$ where ${\mpm(s_i,\act_i,s_{i+1}) > 0}$ for $0 \leq i < n$.
We use $last(\pi)$ to denote the last state of path $\pi$.
Let $Paths^M$ denote the set of all finite paths of~$M$. State $s$ is absorbing if $\supp(\mpm(s, \act)) = \{ s \}$ for all $\act \in \Act$.

\begin{definition}[POMDP]
A \emph{partially observable MDP (POMDP)} is a tuple $\pomdp = \pomdpT$, where $M$ is the underlying MDP, $Z$ is a finite set of observations and $O \colon S \rightarrow Z$ is a (deterministic) observation function. 

\end{definition}
For POMDP $\pomdp$ with underlying MDP $M$, an \emph{observation trace} of path $\pi = s_0\act_0s_1\act_1\dots s_n$ is a sequence $O(\pi) \coloneqq O(s_0)\act_0 O(s_1)\act_1 \dots O(s_n)$. Every MDP can be interpreted as a POMDP with $\Obs = S$ and $O(s) = s$ for all $s \in S$.

A (deterministic) \emph{policy} is a function $\sigma \colon Paths^M \rightarrow \Act$.
Policy $\sigma$ is \emph{memoryless} if $last(\pi) = last(\pi') \Longrightarrow \sigma(\pi) = \sigma(\pi')$ for all $\pi,\pi' \in Paths^M$.
A memoryless policy $\sigma$ maps a state $s \in S$ to action $\sigma(s)$.
Policy $\sigma$ is \emph{observation-based} if $O(\pi) = O(\pi') \Longrightarrow \sigma(\pi) = \sigma(\pi')$ for all $\pi,\pi' \in Paths^M$. For POMDPs, we always consider observation-based policies. We denote by $\Sigma_{\mathit{obs}}$ the set of all observation-based policies. A policy $\sigma \in \Sigma_{\mathit{obs}}$ induces the MC~$\pomdp^{\sigma}$.

We consider indefinite-horizon reachability or expected total reward properties. Formally, let $M=\mdpT$ be an MC, and let $T \subseteq S$ be a set of \emph{target states}.
$\prob[M]{s \models \F{T}}$ denotes the probability of reaching $T$ from state $s \in S$. We use $\prob[M]{\F{T}}$ to denote $\prob[M]{\sinit \models \F{T}}$ and omit the superscript if the MC is clear from context.
Now assume POMDP $\pomdp$ with underlying MDP $M = \mdpT$, and a set $T \subseteq S$ of absorbing target states. Without loss of generality, we assume that the target states are associated with the unique observation $z^T \in Z$, i.e.~$s \in T $ iff $O(s) = z^T$.
For a POMDP $\pomdp$ and $T \subseteq S$, the \emph{maximal reachability probability} of $T$ for state $s \in S$ in $\pomdp$ is $\probmax[\pomdp]{s \models \F{T}} \coloneqq \sup_{\sigma \in \Sigma_{\mathit{obs}}} \statereach{\sigma}{s}$.
The minimal reachability probability $\probmin[\pomdp]{s \models \F{T}} $ is defined analogously.

Finite-state controllers are automata that compactly encode policies. 
\begin{definition}[FSC]
\label{def:fsc}
A \emph{finite-state controller (FSC)} is a tuple $F = \fscT$, with a finite set $N$ of \emph{nodes}, the \emph{initial node} $\ninit \in N$, the \emph{action function} $\gamma \colon N \times Z \rightarrow \Act$ 
and the \emph{update function} $\delta \colon N \times Z \times Z \rightarrow N$.
\end{definition}
A \emph{$k$-FSC} is an FSC with $|N|=k$. If $k{=}1$, the FSC encodes a memoryless policy. 
We use $\ffsc$ ($\ffsck$) to denote the family of all ($k$-)FSCs for POMDP $\pomdp$.
For a POMDP in state $s$, an agent receives observation $\obs = O(s)$.
An agent following an FSC $F$ executes action $\act = \gamma(n,\obs)$ associated with the current node $n$ and the current (prior) observation $z$. 
The POMDP state is updated accordingly to some $s'$ with $\mpm(s, \act, s') > 0$. 
Based on the next (posterior) observation $z' = O(s')$, the FSC evolves to node $n'=\delta(n,z,z')$. 
The \emph{induced MC} for FSC $F$ is $\imc = (S\times N, (\sinit,\ninit), \{\act\}, \mpm^\fsc)$, where for all $(s,n),(s',n') \in S \times N$ we have
\[ \mpm^F \left( (s,n),\act,(s',n') \right) = \iversion{ n' = \delta \left( n,O(s),O(s')\right) } \cdot \mpm(s,\gamma(n,O(s)),s'). \]

We emphasise that for MDPs with infinite state space and POMDPs, an FSC realising the maximal reachability probability generally does not exist. 
For FSC $F \in \ffsc$ with the set $N$ of memory nodes, let $\statereach{F}{(s,n)} \coloneqq \prob[\pomdp^F]{{(s,n) \models \F(T\times N)}}$ denote the probability of reaching target states $T$ from state $(s,n) \in S \times N$. 
Analogously, $\reach{F} \coloneqq \prob[\pomdp^F]{\F{(T\times N)}}$ denotes the probability of reaching target states~$T$ in the MC $\pomdp^F$ induced on $\pomdp$ by $F$.

\medskip
\noindent\textbf{Problem statement.}
 The classical synthesis problem~\cite{madani2005undecidability} for POMDPs asks: given POMDP $\pomdp$, a set $T$ of targets, and a threshold $\lambda$, find an FSC $F$ such that $\reach{F} \geq \lambda$, if one exists. We take a more practical stance and aim instead to optimise the value $\reach{F}$ in an anytime fashion: the faster we can find FSCs with a high value, the better. 
 \begin{remark}
 Variants of the maximising synthesis problem for the expected total reward and minimisation are defined analogously. For conciseness, in this paper, we always assume that we want to maximise the value.
 \end{remark}
In addition to the value of the FSC $F$, another key characteristic of the controller is its \emph{size}, which we treat as a secondary objective and discuss in detail in Sec.~\ref{sec:integrating}.

\section{FSCs for and from Belief Exploration}
\label{sec:beliefbased}
We consider \emph{belief exploration} as described in \cite{bork2022under}. A schematic overview is given in the lower part of Fig.~\ref{fig:saynt}. We recap the key concepts of belief exploration. This section explains two contributions: we discuss how arbitrary FSCs are included and present an approach to export the associated POMDP policies as FSCs.

\subsection{Belief Exploration With Explicit FSC Construction}
\label{subsec:belief}

Finite-state controllers for a POMDP can be obtained by analysing the (fully observable) \emph{belief MDP}~\cite{smallwood1973optimal}. The state space of this MDP consists of \emph{beliefs}: probability distributions over states of the POMDP $\pomdp$ having the same observation. Let $S_\obs \coloneqq \{s \in S \mid O(s) = \obs\}$ denote the set of all states of $\pomdp$ with observation $\obs \in \Obs$. Let the set of all beliefs $\beliefs \coloneqq \bigcup_{z \in Z} Distr(S_\obs)$ and denote for $b \in \beliefs$ by $O(b) \in Z$ the unique observation $O(s)$ of any $s \in \supp(b)$.

In a belief $b$, taking action $\act$ yields an updated belief as follows: let $\mpm(b,\act,z') \coloneqq \sum_{s \in S_{O(b)}}b(s) \cdot \sum_{s'\in S_{z'}}\mpm(s,\act,s')$ denote the probability of observing $z' \in Z$ upon taking action $\act \in \Act$ in belief $b \in \beliefs$.
If $\mpm(b,\act,z')>0$, the corresponding successor belief $b'=\bsucc{b}{\act}{z'}$ with $O(b')=z'$ is defined component-wise as

$$ \bsucc{b}{\act}{z'}(s') \coloneqq \frac{\sum_{s \in S_{O(b)}}b(s)\cdot \mpm(s,\act,s')}{\mpm(b,\act,z')} $$
for all $s' \in S_{z'}$. Otherwise, $\bsucc{b}{\act}{z'}$ is undefined.

\begin{definition}[Belief MDP] The \emph{belief MDP} of POMDP $\pomdp$ is the MDP $\bmdp = \bmdpT$, with initial belief $\binit \coloneqq \{\sinit \mapsto 1\}$ and transition function $\bpm(b,\act,b') \coloneqq \iversion{ b'=\bsucc{b}{\act}{z'} } \cdot \mpm(b,\act,z')$ where $z'=O(b')$.
\end{definition}
The belief MDP captures the behaviour of its POMDP. It can be unfolded by starting in the initial belief and computing all successor beliefs.

\paragraph{Deriving FSCs from finite belief MDPs.}
Let $\btarget \coloneqq \left\{ b \in \beliefs \mid O(b) = z^T\right\}$ denote the set of \emph{target beliefs}.
If the reachable state space of the belief MDP $\bmdp$ is finite, e.g. because the POMDP is acyclic, standard model checking techniques can be applied to compute the memoryless policy $\bsigma \colon \beliefs \rightarrow \Act$ that selects in each belief state ${b \in \beliefs}$ the action that maximises $\prob{b \models \F{\btarget}}$\footnote{Memoryless policies suffice to maximise the value in a fully observable MDP~\cite{puterman1994}.}.
We can translate the deterministic, memoryless policy $\bsigma$ into the corresponding FSC $\fscbel = \left(\beliefs, \binit, \gamma, \delta \right)$ with action function $\gamma(b, z) = \bsigma(b)$ and update function $\delta(b, z, z') = \bsucc{b}{\bsigma(b)}{z'}$ for all $z,z' \in Z$.\footnote{The assignments of missing combinations where $z \neq O(b)$ are irrelevant.}

\paragraph{Handling large and infinite belief MDPs.}
In case the reachable state space of the belief MDP $\bmdp$ is infinite or too large for a complete unfolding, a finite approximation $\clipmdp$ is used instead~\cite{bork2022under}. Assuming $\bmdp$ is unfolded up to some depth, let $\exploredbel \subset \beliefs$ denote the set of explored beliefs and let $\frontier \subset \beliefs {\setminus} \exploredbel$ denote the \emph{frontier}: the set of unexplored beliefs reachable from $\exploredbel$ in one step.
To complete the finite abstraction, we require handling of the frontier beliefs. The idea is to use for each $b \in \mathcal{U}$ a
\emph{cut-off value} $\lb(b)$: an under-approximation of the maximal reachability probability $\probmax[\bmdp]{b \models \F{\btarget}}$ for $b$ in the belief MDP. We explain how to compute cut-off values systematically given an FSC in Sec.~\ref{subsec:paynt-to-storm}.

Ultimately, we define a finite MDP $\clipmdp = (\exploredbel \cup \frontier \cup \{\btop,\bbot\}, \binit, \Act, \overline{\bpm})$ with the transition function: $\overline{\bpm}(b,\alpha) \coloneqq \bpm(b,\act)$ for explored beliefs $b \in \mathcal{E}$ and all $\act \in \Act$, and
$\overline{\bpm}(b,\act) \coloneqq \{\btop \mapsto \lb(b), \bbot \mapsto 1-\lb(b)\}$ for frontier beliefs $b \in \frontier$ and all $\act \in \Act$, where $\btop$ and $\bbot$ are fresh sink states, i.e.~$\overline{\bpm}(\btop,\act) \coloneqq \{\btop \mapsto 1\}$ and $\overline{\bpm}(\bbot,\act) \coloneqq \{\bbot \mapsto 1\}$ for all $\act \in \Act$.
The reachable state space of $\clipmdp$ is finite, enabling its automated analysis; since our method to compute cut-off values emulates an FSC, a policy maximising $\probmax[\clipmdp]{\F{(\btarget \cup \{\btop\})}}$ induces an FSC for the original POMDP $\pomdp$. We discuss how to obtain this FSC in Sec.~\ref{subsec:belieffsc}.

\subsection{Using FSCs for Cut-off Values}
\label{subsec:paynt-to-storm}

A crucial aspect when applying the belief exploration with cut-offs is the choice of suitable cut-off values. The closer the cut-off value is to the actual optimum in a belief, the better the approximation we obtain. In particular, if the cut-off values coincide with the optimal value, cutting off the initial state is optimal.
However, finding optimal values is as hard as solving the original POMDP. We consider \emph{under-approximative value functions} induced by applying \emph{any}\footnote{We remark that~\cite{bork2022under} considers memoryless FSCs only.} FSC to the POMDP and lifting the results to the belief MDP. The better the FSC, the better the cut-off value. 
We generalise belief exploration with cut-offs such that the approach supports arbitrary sets of FSCs with additional flexibility.

Let $\fscind \in \ffsc$ be an arbitrary, but fixed FSC for POMDP $\pomdp$. Let $p_{s,n} \coloneqq \prob[\pomdp^{\fscind}]{(s,n) \models \F{T}}$ for state $(s,n) \in S \times N$ in the corresponding induced MC.
For fixed $n \in N$, $V(b,n) \coloneqq \sum_{s \in S_{O(b)}} b(s) \cdot p_{s,n}$ denotes the cut-off value for belief $b$ and memory node $n$. It corresponds to the probability of reaching a target state in $\pomdp^{\fscind}$ when starting in memory node $n \in N$ and state $s \in S$ according to the probability distribution $b$.
We define the overall cut-off value for $b$ induced by $F$ as $\lb(b) \coloneqq \max_{n \in N} V(b,n)$. It follows straightforwardly that $\lb(b) \leq \probmax[\bmdp]{b \models \F{\btarget}}$.
As values $p_{s,n}$ only need to be computed once, computing $\lb(b)$ for a given belief $b$ is relatively simple. However, the complexity of the FSC-based cut-off approach depends on the size of the induced MC. Therefore, it is essential that the FSCs used to compute cut-off values are concise.

\subsection{Extracting FSC from Belief Exploration}
\label{subsec:belieffsc}
Model checking the finite approximation MDP $\clipmdp$ with cut-off values induced by an FSC $\fscind$ yields a maximising memoryless policy $\bsigma$. 
Our goal is to represent this policy as an FSC $\fscbel$. 
We construct $\fscbel$ by considering both $\fscind$ and the necessary memory nodes for each explored belief $b \in \exploredbel$. 
Concretely, for each explored belief, we introduce a corresponding memory node. 
In each such node, the action $\bsigma(b)$ is selected. 
For the memory update, we distinguish between two cases based on the next belief after executing $\bsigma(b)$ in $\clipmdp$. 
If for observation $\obs' \in \Obs$, the successor belief $b' = \bsucc{b}{\bsigma(b)}{z'} \in \exploredbel$, the memory is updated to the corresponding node. 
Otherwise, $b' \in \frontier$ holds, i.e., the successor is part of the frontier. The memory is then updated to the memory node $n$ of FSC $\fscind$ that maximises the cut-off value $V(b',n)$. This corresponds to the notion that if the frontier is encountered, we switch from acting according to policy $\bsigma$ to following $\fscind$ (initialised in the correct memory node). This is formalised as:

\begin{definition}[Belief-based FSC with cut-offs]
Let $\fscind = (N,n_0,\gamma_{\inductive}, \delta_{\inductive})$ and $\clipmdp$ as before.
The \emph{belief-based FSC with cut-offs} is $\fscbel = (\exploredbel \cup N, \binit, \gamma, \delta)$ with action function $\gamma(b,z) = \bsigma(b)$ for $b \in \exploredbel$ and $\gamma(n,z) = \gamma_{\inductive}(n,z)$ for $n \in N$ and arbitrary $z \in Z$. The update function $\delta$ is defined for all $z, z' \in Z$ by $\delta(n,z,z') = \delta_{\inductive}(n,z,z')$ if $n \in N$, and for $b \in \exploredbel$ with $b' = \bsucc{b}{\bsigma(b)}{z'}$ by:
$$
\delta(b,z,z') = b' \mbox{ if } b' \in \exploredbel, \mbox{ and } \delta(b,z,z') = \mathrm{argmax}_{n \in N}V(b',n) \mbox{ otherwise.} 
$$
\vspace*{-0.5cm}
\end{definition} 

\vspace{-1.5em}
\section{Accelerated Inductive Synthesis}
\label{sec:synthesis}

In this section, we consider inductive synthesis~\cite{andriushchenko2022inductive}, an approach for finding controllers for POMDPs in a set of FSCs.
We briefly recap the main idea, then first explain how to use a reference policy. Finally, we introduce and discuss a novel search space for the controllers that we consider in this paper in detail.

\subsection{Inductive Synthesis with $k$-FSCs}

In the scope of this paper, inductive synthesis~\cite{andriushchenko2021inductive} considers a finite family of FSCs $\ffsck$ of $k$-FSCs with memory nodes $N = \{n_0,\dots,n_{k-1}\}$, and the family $\fimc \coloneqq \{\imc \mid F \in \ffsck\}$ of associated induced MCs. 
The states for each MC are tuples $(s, n) \in S \times N$.
For conciseness, we only discuss the abstraction-refinement framework~\cite{ceska2019shepherding} within the inductive synthesis loop. The overall image is as in Fig.~\ref{fig:saynt}. 
Informally, the \emph{MDP abstraction} of the family $\fimc$ of MCs is an MDP $\fmdp$ with the set $S \times N$ of states such that, if some MC $M \in \fimc$ executes action $\act$ in state $(s,n) \in S \times N$, then this action (with the same effect) is also enabled in state $(s,n)$ of $\fmdp$. Essentially, $\fmdp$ over-approximates the behaviour of all the MCs in the family $\fimc$: it simulates an arbitrary family member in every step, but it may switch between steps.\footnote{The MDP is an game-based abstraction~\cite{DBLP:conf/qest/KwiatkowskaNP06} of the all-in-one MC~\cite{DBLP:journals/fac/ChrszonDKB18}.}
\begin{definition}
\label{def:mdpapproxnaive}
\emph{MDP abstraction} for POMDP $\pomdp$ and family $\ffsck = \{ \fsc_1, \hdots, \fsc_m \}$ of $k$-FSCs is the MDP $\fmdp \coloneqq \big(S\times N, (\sinit,\ninit), \{ 1, \hdots, m \}, \mpm^\ffsck \big)$
with 
\[ \mpm^\ffsck ( (s,n), i ) = \mpm^{F_i}. \]
\end{definition}
While this MDP has $m$ actions, practically, many actions coincide. Below, we see how to utilise the structure of the FSCs. Here, we finish by observing that the MDP is a proper abstraction:
\begin{lemma}\cite{ceska2019shepherding}
For all $\fsc \in \ffsck$, $\probmin[\fmdp]{\F{T}} \leq \prob[\imc]{\F{T}} \leq \probmax[\fmdp]{\F{T}}$.
\end{lemma}
With that result, we can naturally start with the set of all $k$-FSCs and search through this family by selecting suitable subsets~\cite{ceska2019shepherding}.
Since the number $k$ of memory nodes necessary is not known in advance, one can iteratively explore the sequence $\ffsc_{1},\ffsc_{2},\dots$ of families of FSCs of increasing complexity.

\subsection{Using Reference Policies to Accelerate Inductive Synthesis}
\label{subsec:storm-to-paynt}
Consider the synthesis process of the optimal $k$-FSC $F \in \ffsck$ for POMDP $\pomdp$.
To accelerate the search for $F$ within this family, we consider a reference policy, e.g., a policy $\bsigma$ extracted from an (approximation of the) belief MDP, and  shrink the FSC family. 
For each observation $z \in Z$, we collect the set $\Act[\bsigma](z) \coloneqq \left\{ \bsigma(b) \mid b \in \beliefs, O(b) = z \right\}$ of actions that were selected by $\bsigma$ in beliefs with observation $z$.
The set $\Act[\bsigma](z)$ contains the actions used by the reference policy when in observation $z$.
We focus the search on these actions by constructing a subset of FSCs $\{\ (N, \ninit, \gamma, \delta) \in \ffsck \mid \forall n \in N, \obs \in \Obs. \gamma(n, z) \in \Act[\bsigma](z) \}$.

Restricting the action selection may exclude the optimal $k$-FSC. It also does not guarantee that the optimal FSC in the restricted family achieves the same value as the reference policy $\bsigma$ as $\bsigma$ may have more memory nodes. 
We first search the restricted space of FSCs before searching the complete space. This also accelerates the search: The earlier a good policy is found, the easier it is to discard other candidates (because they are provably not optimal). Furthermore, in case the algorithm terminates earlier (notice the anytime aspect of our problem statement), we are more likely to have found a reasonable policy. 

Additionally, we could use sets $\Act[\bsigma]$ to determine with which $k$ to search. If in some observation $z \in Z$ the belief policy $\bsigma$ uses $|\Act[\bsigma](z)|$ distinct actions,
then in order to enable the use of all of these actions, we require at least $k = \max_{\obs \in \Obs} |\Act[\bsigma](z)|$ memory states. However, this may lead to families that are too large and thus we use a more refined view discussed below.

\subsection{Inductive Synthesis with Adequate FSCs}
In this section, we discuss the set of candidate FSCs in more detail. In particular, we take a more refined look at the families that we consider.

\paragraph{More granular FSCs.}
We consider memory models~\cite{andriushchenko2022inductive} that describe per-observation how much memory may be used:
\begin{definition} [$\mu$-FSC]
A \emph{memory model} for POMDP $\pomdp$ is a function $\mu \colon Z \rightarrow \naturals$. Let $k = \max_{z \in Z}\mu(z)$. The $k$-FSC $F \in \ffsck$ with nodes $N = \{n_0,\dots,n_{k-1}\}$ is a \emph{$\mu$-FSC} iff for all $z \in Z$ and for all $i > \mu(z)$ it holds: $\gamma(n_i,z) = \gamma(\ninit,z)$ and $\delta(n_i,z,z') = \delta(\ninit,z,z')$ for any $z' \in Z$.

\end{definition}
 $\ffscmu$~denotes the family of all $\mu$-FSCs.
Essentially, memory model $\mu$ dictates that for prior observation $z$ only $\mu(z)$ memory nodes are utilised, while the rest behave exactly as the default memory node $\ninit$.
Using memory model $\mu$ with $\mu(z)<k$ for some observations $z \in Z$ greatly reduces the number of candidate controllers.
For example, if $|S_\obs|=1$ for some $\obs \in \Obs$, then upon reaching this state, the history becomes irrelevant. It is thus sufficient to set $\mu(z)=1$ (for the specifications in this paper).
It also significantly reduces the size of the abstraction, see Appendix~A of~\cite{andriushchenko2023search}.

\begin{figure}[t]
\centering
\begin{subfigure}{.65\textwidth}
\centering
\includegraphics[width=0.8\linewidth]{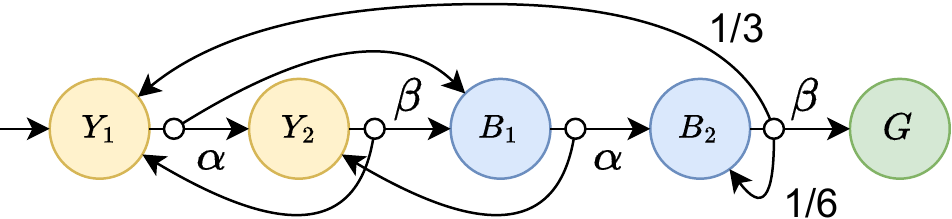}
\caption{}
\label{fig:aware-pomdp}
\end{subfigure}%
\begin{subfigure}{.3\textwidth}
\centering
\includegraphics[width=0.9\linewidth]{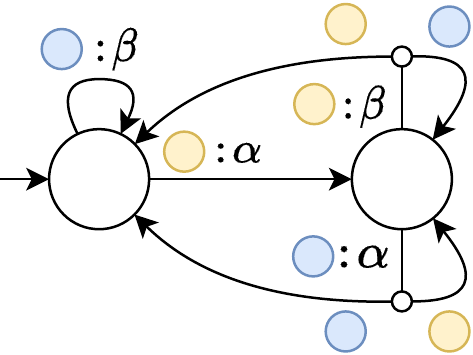}
\caption{}
\label{fig:aware-fsc}
\end{subfigure}%
\vspace{-0.6em}
\caption{(a) A POMDP where colours and capital letters encode observations; unlabelled transitions have probability $1/2$; omitted actions (e.g. action $\beta$ in the initial state) are self-loops; the objective is to minimise the expected number of steps to reach state $G$.
(b)~The optimal posterior-aware 2-FSC.}
\vspace{-.5em}
\label{fig:aware}
\end{figure}

\paragraph{Posterior-aware or posterior-unaware.}
The technique outlined in~\cite{andriushchenko2022inductive} considers \emph{posterior-unaware FSCs}~\cite{DBLP:conf/aaai/AmatoBZ10}.
An FSC with update function $\delta$ is posterior-unaware if the posterior observation is not taken into account when updating the memory node of the FSC, i.e.~ $\delta(n,z,z') = \delta(n,z,z'')$ for all $n\in N, z, z',z'' \in Z$.
This restriction reduces the policy space and thus the MDP abstraction $\fmdp$.
On the other hand, general (posterior-aware) FSCs can utilise information about the next observation to make an informed decision about the next memory node.
As a result, fewer memory nodes are needed to encode complex policies.
Consider Fig.~\ref{fig:aware-pomdp} which depicts a simple POMDP.
First, notice that in yellow states $Y_i$ we want to be able to execute two different actions, implying that we need at least two memory nodes to distinguish between the two states, and the same is true for the blue states $B_i$.
Second, notice that in each state the visible action always leads to states having different observations, implying that the posterior observation $z'$ is crucial for the optimal decision making. If $z'$ is ignored, it is impossible to optimally update the memory node.
Figure~\ref{fig:aware-fsc} depicts the optimal posterior-aware 2-FSC allowing to reach the target within 12 steps on expectation.
The optimal posterior-unaware FSC has at least 4 memory nodes and the optimal posterior-unaware 2-FSC uses 14 steps.

\paragraph{MDP abstraction.}
To efficiently and precisely create and analyse MDP abstractions, Def.~\ref{def:mdpapproxnaive} is overly simplified. In Appendix~A of~\cite{andriushchenko2023search}, we present the construction for general, posterior-aware FSCs including memory models.

\section{Integrating Belief Exploration with Inductive Synthesis}
\label{sec:integrating}
\newcommand{\method}[1]{\texttt{\textcolor{blue!50!black}{#1}}}
\newcommand{\searchmethod}{\method{search}}
\newcommand{\explmethod}{\method{explore}}

We clarify the symbiotic approach from Fig.~\ref{fig:saynt} and review FSC sizes.

\subsubsection{Symbiosis by closing the loop}
Section~\ref{sec:beliefbased} shows the potential to improve belief exploration using FSCs, e.g., obtained from an inductive synthesis loop, whereas Sec.~\ref{sec:synthesis} shows the potential to improve inductive synthesis using policies from, e.g., belief exploration. 
A natural next step is to use improved inductive synthesis for belief exploration and improved belief exploration for inductive synthesis, i.e., to alternate between both techniques. This section briefly clarifies the symbiotic approach from Fig.~\ref{fig:saynt} using Alg.~\ref{alg:integrated}.
\begin{algorithm}[t]
\SetKwInOut{Input}{Input}
\SetKwInOut{Output}{Output}
\SetKw{Continue}{continue}
\SetKw{Yield}{yield}
\SetKwComment{Comment}{$\triangleright$\ }{}
\Input{POMDP $\pomdp$, set $T$ of target states, timeout values $t, t_{\inductive}, t_{\belief}$}
\Output{Best FSCs $\fscind$ and $\fscbel$ found so far}

\BlankLine
$\fscind \gets \bot$, $\mathcal{F} \gets \mathcal{F}_1^\pomdp$, $k \gets 0$, $\mu \gets \{z \mapsto 1 \mid z \in Z \}$, 
$\fscbel \gets \bot$, $\bsigma \gets \bot$ \\ 
\While{\textrm{not timeout} $t$} {
    \While{\textrm{not timeout} $t_{\inductive}$} { 
    \If{$\mathcal{F} = \emptyset$} {
        $k \gets k+1$ \\
        $\forall z \in Z \colon \mu(z) \gets \max \{ \mu(z), k \}$ \\
        $\mathcal{F} \gets \ffscmu$ \\
    }
    $\mathcal{F},\fscind \gets \searchmethod(\mathcal{F},\fscind,\Act[\bsigma] \textbf{ if } \reach{\fscind} > \reach{\fscbel} \textbf{ else } \bot )$\\
}

$\bsigma,\fscbel \gets \explmethod(t_{\belief}, \fscind)$\\
\If{$\reach{\fscind} \leq \reach{\fscbel}$ \text{ and } $\exists z \in Z \colon \mu(z) < |\Act[\bsigma](z)|$} {
$\forall z \in Z \colon \mu(z) \gets |\Act[\bsigma](z)| $ \\
$\mathcal{F} \gets \ffscmu$ \\
}
\Yield{$\fscind, \fscbel$}
}
\caption{Anytime algorithm}
\label{alg:integrated}
\end{algorithm}

We iterate until a global timeout $t$: 
in each iteration, we make both controllers available to the user as soon as they are computed (Alg.~\ref{alg:integrated}, l.~13).
We start in the inductive mode (l.~3-8), where we initially consider the $1$-FSCs represented in $\ffscmu$. 
Method $\searchmethod$ (l.~8) investigates $\mathcal{F}$ and outputs the new maximising FSC \fscind (if it exists). 
If the timeout $t_{\inductive}$ interrupts the synthesis process, the method additionally returns yet unexplored parameter assignments.
If $\mathcal{F}$ is fully explored within the timeout $t_{\inductive}$ (l.~4), we increase $k$ and repeat the process.
After the timeout~$t_{\inductive}$, we run belief exploration $\explmethod$ for $t_{\belief}$ seconds, where we use $\fscind$ as backup controllers (l.~9). 
After the timeout~$t_{\belief}$ (exploration will continue from a stored configuration in the next belief phase), we use \fscind to obtain cut-off values at unexplored states, compute the optimal policy $\bsched$ (see Sec.~\ref{sec:beliefbased}) and extract the FSC \fscbel which incorporates \fscind.
Before we continue the search, we check whether the belief-based FSC is better and whether that FSC gives any reason to update the memory model (l.~10). If so, we update $\mu$ and reset the $\mathcal{F}$ (l.~11-12).

\subsubsection{The size of an FSC}
We have considered several sub-classes of FSCs and wish to compare the sizes of these controllers. For FSC $F = \fscT$, we define its size $\size(F) \coloneqq \size(\gamma) +\size(\delta)$ as the memory required to encode functions $\gamma$ and $\delta$.
Encoding  $\gamma \colon N \times Z \rightarrow Act$ of a general $k$-FSC requires $\size(\gamma) = \sum_{n \in N} \sum_{z \in Z} 1 = k \cdot |Z|$ memory. 
Encoding $\delta \colon N \times Z \times Z \rightarrow N$ requires $k \cdot |Z|^2$ memory.
However, it is uncommon that in each state-memory pair $(s,n)$ all posterior observations can be observed.
We therefore encode $\delta(n,z,\cdot)$ as a sparse adjacency list, i.e.,~as a list of pairs $(z',\delta(n,z,z'))$.
To define the size of such a list properly, consider the induced MC $\pomdp^F = (S \times N, (\sinit,\ninit), \{\act\}, \mpm^F)$.
Let $post(n,z) \coloneqq \left\{O(s') \mid \exists s \in S_z \colon (s',\cdot) \in \supp(\mpm^F((s,n),\act)) \right\}$ denote the set of posterior observations reachable when taking a transition in a state $(s,n)$ of $\pomdp^F$ with $O(s) = z$.
Table~\ref{tab:size-formulae} summarises the  resulting sizes of FSCs of various sub-classes. The derivation is included 
in Appendix~B of~\cite{andriushchenko2023search}.
Table~\ref{tab:size} on p.~\pageref{tab:size} shows that we typically find much smaller $\mu$-FSCs ($\fscind$) than belief-based FSCs~($\fscbel$).

\begin{table*}[t]
\centering
\renewcommand{\arraystretch}{1.3}
\begin{tabular}{|c|c|c|}
\hline
FSC class & $\size(\gamma)$ & $\size(\delta)$ \\ \hline
$k$-FSC & $k \cdot |Z|$ & $2 \cdot \sum_{n \in N} \sum_{z \in Z} |post(n,z)|$ \\ 
$\mu$-FSC & $\sum_{z \in Z} \mu(z)$ & $2 \cdot \sum_{z \in Z} \sum_{i=0}^{\mu(z)-1} |post(n_i,z)|$ \\
posterior-unaware $\mu$-FSC & $\sum_{z \in Z} \mu(z)$ & $\sum_{z \in Z} \mu(z)$\\
$\fscbel$ using $\fscind$ for cut-offs & $\size(\gamma_{\inductive}) + |\exploredbel|$ & $\size(\delta_{\inductive}) + 2 \cdot \sum_{b \in \exploredbel} |post(b,O(b))|$ \\
\hline
\end{tabular}

\vspace{0.5em}
\caption{Sizes of different types of FSCs.}
\vspace{-1.5em}
\label{tab:size-formulae}
\end{table*}

\section{Experiments}

\label{sec:experiments}

Our evaluation focuses on the following three questions:
\begin{itemize}
\item[Q1:] \emph{Do the FSCs from inductive synthesis raise the accuracy of the belief MDP?}

\noindent
\item[Q2:] \emph{Does exploiting the belief MDP boost the inductive synthesis of FSCs?}

\noindent
\item[Q3:] \emph{Is the symbiotic approach improving run time, controller's value and size?}
\end{itemize}

\subsubsection{Selected benchmarks and setup}
Our baseline are the recent belief exploration technique~\cite{bork2022under} implemented in \storm~\cite{STORM} and the inductive (policy) synthesis method~\cite{andriushchenko2022inductive} implemented in \paynt~\cite{andriushchenko2021paynt}. 
\paynt uses \storm for parsing and model checking of MDPs, but not for solving POMDPs. 
Our symbiotic framework (Alg.~\ref{alg:integrated}) has been implemented on top of \paynt and \storm. 
In the following, we use \storm and \paynt to refer to the implementation of belief exploration and inductive synthesis respectively, and \integration to refer to the symbiotic framework. The implementation of \integration and all benchmarks are publicly available\footnote{
\url{https://github.com/randriu/synthesis}}. Additionally, the implementation and the benchmarks in the form of an artifact are also available at \url{https://doi.org/10.5281/zenodo.7874513}.

\paragraph{Setup.} The experiments are run on a single core of a machine equipped with an Intel i5-12600KF @4.9GHz CPU and 64GB of RAM. \paynt searches for posterior-unaware FSCs using abstraction-refinement, as suggested by~\cite{andriushchenko2022inductive}. 
By default, \storm applies the cut-offs as presented in Sect.~\ref{subsec:belief}.
\integration uses the default settings for \paynt and \storm while $t_{\inductive} = 60s$ and $t_{\belief} = 10s$ were taken for Alg.~\ref{alg:integrated}.
Under Q3, we discuss the effect of changing these values.
 
\paragraph{Benchmarks.}
We evaluate the methods on a selection of models from \cite{bork2020verification,bork2022under,andriushchenko2022inductive} supplemented by larger variants of these models (Drone-8-2 and \mbox{Refuel-20}), by one model from~\cite{hauskrecht1997incremental} (Milos-97) and by the synthetic model (Lanes+) described in Appendix~C of~\cite{andriushchenko2023search}. We excluded benchmarks for which \paynt or \storm finds the (expected) optimal solution in a matter of seconds. The benchmarks were selected to illustrate advantages as well as drawbacks of all three synthesis approaches: belief exploration,  inductive (policy) search, and the symbiotic technique. 
Table~\ref{tab:stats} lists for each POMDP the number $|S|$ of states, the total number $\sum \Act \coloneqq \sum_{s}|\Act(s)|$ of actions, the number $|Z|$ of observations, the specification (either maximising or minimising a reachability probability $P$ or expected reward $R$), and a known over-approximation on the optimal value computed using the technique from~\cite{bork2020verification}. These over-approximations are solely used as rough estimates of the optimal values.
Tab.~\ref{tab:integration} on p.~\pageref{tab:integration} reports the quality of the resulting FSCs on a broader range of benchmarks and demonstrates the impact of the non-default settings.

\begin{table}[t]

\renewcommand{\arraystretch}{0.95}
\setlength{\tabcolsep}{2pt}
\scalebox{0.94}{
\begin{tabular}{|c|ccc|r|c||c|ccc|r|c|}
\hline
\multirow{2}{*}{Model} & \multirow{2}{*}{$|S|$} & \multirow{2}{*}{${\sum}\Act$} & \multirow{2}{*}{$|Z|$} & \multirow{2}{*}{Spec.} & Over- & \multirow{2}{*}{Model} & \multirow{2}{*}{$|S|$} & \multirow{2}{*}{${\sum}\Act$} & \multirow{2}{*}{$|Z|$} & \multirow{2}{*}{Spec.} & Over-\\
& & & & & approx. & & & & & & approx.
\\ \hline 

4x3-95 & 22 & 82 & 9 & $R_{\max}$ & $\leq 2.24$ & 
Drone-4-2 & 1226 & 2954 & 761 & $P_{\max}$ & $\leq 0.98$ 
\\ 

4x5x2-95 & 79 & 310 & 7 & $R_{\max}$ & $\leq 3.26$ &
Drone-8-2 & 13k & 32k & 3195 & $P_{\max}$ & $\leq 0.99$ \\

Hallway & 61 & 301 & 23 & $R_{\min}$ & $\geq 11.5$ & 
Lanes+ & 2741 & 5285 & 11 & $R_{\min}$ & $\geq 4805$
 \\ 

Milos-97 & 165 & 980 & 11 & $R_{\max}$ & $\leq 80$ & 
Netw-3-8-20 & 17k & 30k & 2205 & $R_{\min}$ & $\geq 4.31$

\\ 

 Network & 19 & 70 & 5 & $R_{\max}$ & $\leq 359$ & 
 Refuel-06 & 208 & 565 & 50 & $P_{\max}$ & $\leq 0.78$

\\ 

Query-s3 & 108 & 320 & 6 & $R_{\max}$ & $\leq 600$ & 
 Refuel-20 & 6834 & 25k & 174 & $P_{\max}$ & $\leq 0.99$

\\ 

Tiger-95 & 14 & 50 & 7 & $R_{\max}$ & $\leq 159$ & 
Rocks-12 & 6553 & 32k & 1645 & $R_{\min}$ & $\geq 17.8$

\\ \hline
\end{tabular}
}
\vspace{0.5em}
\caption{Information about the benchmark POMDPs.}

\label{tab:stats}
\vspace{-1.5em}
\end{table}

\subsection*{Q1: FSCs provide better approximations of the belief MDP}

In these experiments, \paynt is used to obtain a sub-optimal \fscind within 10s  which is then used by \storm.
Tab.~\ref{tab:oneway} (left) lists the results.
Our main finding is that \textcolor{blue}{\emph{belief exploration can yield better FSCs (and sometimes faster) using FSCs from \paynt}---even if the latter FSCs are far from optimal}.
For instance, \storm with provided $\fscind$ finds an FSC with value 0.97 for the Drone-4-2 benchmark within a total of 10s (1s+9s for obtaining \fscind), compared to obtaining an FSC of value 0.95 in 56s on its own.
A value improvement is also obtained if \storm runs longer.
For the Network model, the value improves with 37\% (short-term) and 47\% (long-term) respectively, at the expense of investing 3s to find \fscind.
For the other models, the relative improvement ranges from 3\% to 25\%.
A further value improvement can be achieved when using better FSCs \fscind from \paynt; see Q3.
Sometimes, belief exploration does not profit from \fscind.
For Hallway, the unexplored part of the belief MDP becomes insignificant rather quickly, and so does the impact of \fscind.
Clipping~\cite{bork2022under}, a computationally expensive extension of cut-offs, is beneficial only for Rocks-12, rendering \fscind useless. 
Though even in this case, using \fscind significantly improves Short \storm that did not have enough time to apply~clipping.

\begin{table*}[t]
\centering
 \setlength{\tabcolsep}{1pt}

\begin{tabular}{|c||r||r|r||r|r|}
\hline

& 
\multicolumn{1}{c||}{\paynt} & 
\multicolumn{2}{c||}{Short \storm} & 
\multicolumn{2}{c|}{Long \storm}\\

\multicolumn{1}{|c||}{Model} & 
\multicolumn{1}{c||}{\fscind}& 
\multicolumn{1}{c|}{} & \multicolumn{1}{c||}{+ \fscind} & 
\multicolumn{1}{c|}{} & \multicolumn{1}{c|}{+\fscind}\\ \hline \hline

Drone-4-2 & 
$0.94$ & 
$0.92$ & 
${0.97}$ &
$0.95$ &
$0.97$ \\ 

$P_\mathrm{max}$ & 
$9s$&
$1s$ &
$1s$ &
$56s$&
$57s$
\\ \hline

Network & 
$266.1$ & 
$186.7$ & 
$274.5$ &
$202.1$&
${277.1}$ \\ 

$R_\mathrm{max}$&
$3s$&
${<}1s$ &
${<}1s$ &
$26s$&
$33s$
\\ \hline

Drone-8-2 & 
$0.9$ & 
$0.6$ & 
${0.96}$ &
$0.68$ &
$0.97$ \\ 

$P_\mathrm{max}$ & 
$28s$&
$3s$ &
$3s$ &
$101s$&
$103s$
\\ \hline

4x3-95 & 
$1.66$ & 
$1.62$ & 
$1.82$ &
$1.84$ &
${1.88}$ 
\\ 

$R_\mathrm{max}$ & 
$7s$&
${<}1s$ &
${<}1s$ &
$60s$&
$72s$
\\ \hline

Query-s3 & 
$425.2$ & 
$417.4$ & 
$430.0$ &
$419.6$&
${432.0}$ \\ 

$R_\mathrm{max}$&  
$7s$&
$2s$ &
$2s$ &
$91s$&
$94s$
\\ \hline

Milos-97& 
$31.56$ & 
$37.15$ & 
$39.15$ &
$38.35$&
${40.64}$ \\ 
 
$R_\mathrm{max}$& 
$3s$&
${<}1s$ &
${<}1s$ &
$42s$&
$42s$
\\ \hline

Hallway & 
$16.05$ & 
$13.07$ & 
$12.63$ &
$12.55$&
$12.55$ \\ 

$R_\mathrm{min}$& 
$9s$&
$1s$ &
$1s$ &
$160s$&
$167s$
\\ \hline

Rocks-12& 
$42$ & 
$38$ & 
$31.89$ &
$20$*&
$20$*\\ 

$R_\mathrm{min}$&  
${<}1s$&
${<}1s$ &
${<}1s$ &
$10s$&
$10s$
\\ \hline

\end{tabular}
\hspace{0.6em}
\begin{tabular}{|c||r||r|r|}
\hline

\renewcommand{\arraystretch}{0.95}
 \setlength{\tabcolsep}{1pt}

& 
\multicolumn{1}{c||}{\storm{}} &  
\multicolumn{2}{c|}{\paynt}\\

\multicolumn{1}{|c||}{Model} &  
\multicolumn{1}{c||}{\fscbel}& 
\multicolumn{1}{c|}{} & \multicolumn{1}{c|}{+ \fscbel}\\ \hline \hline

4x5x2-95& 
$2.08$ & 
$0.94$ &
${2.03}$ \\

$R_\mathrm{max}$&  
${<}1s$&
$258s$ &
$38s$
\\ \hline

Refuel-20& 
$0.09$ & 
${<}0.01$ &
$0.19$  \\ 

$P_\mathrm{max}$& 
$1s$ &
$10s$ &
$11s$ 
\\ \hline

Tiger-95& 
$50.38$ & 
$2.99$ &
${28.73}$  \\ 

$R_\mathrm{max}$ & 
${<}1s$ &
$14s$ &
$23s$
\\ \hline

4x3-95& 
$1.62$ &  
$1.75$ &
${1.84}$  \\ 

$R_\mathrm{max}$& 
${<}1s$ &
$14s$ &
$238s$ 
\\ \hline

Refuel-06 & 
$0.67$ & 
$0.35$ &
${0.67}$ \\ 

$P_\mathrm{max}$ & 
${<}1s$ &
${<}1s$ &
$42s$ 
\\ \hline

Milos-97& 
$37.15$ & 
$31.56$ &
${39.29}$ \\

$R_\mathrm{max}$ &  
${<}1s$ &
$3s$ &
$215s$
\\ \hline

Netw-3-8-20& 
$11.93$ & 
$11.07$ &
${10.95}$  \\ 

 $R_\mathrm{min}$ &  
$1s$ &
$185s$ &
$271s$ 
\\ \hline


Rocks-12 & 
$38$ & 
$42$ &
${38}$ \\ 

$R_\mathrm{min}$ &  
${<}1s$ &
${<}1s$ &
${<}1s$ 

\\ \hline

\end{tabular}

\vspace{0.5em}
\caption{
\textbf{Left (Q1)}: 
Experimental results on how a (quite sub-optimal) FSC \fscind computed by \paynt within 10s impacts \storm. (For Drone-8-2, the largest model in our benchmark, we use 30s).
The ``\paynt{}" column indicates the value of \fscind and its run time.
The ``Short \storm{}" column runs storm for 1s and compares the value of FSC \fscbel found by \storm alone to \storm using \fscind.
The ``Long \storm{}" column is analogous, but with a 300s timeout for \storm. 
In the last row, * indicates that clipping was used. 
\textbf{Right (Q2)}: 
Experimental results on how an FSC \fscbel obtained by a shallow exploration of the belief MDP impacts the inductive synthesis by \paynt. 
The ``\storm{}" column reports the value of \fscbel computed within 1s. 
The ``\paynt{}" column compares the values of the FSCs \fscind obtained by \paynt itself to \paynt using the FSCs \fscbel within a 300s timeout.
} 
\label{tab:oneway}
\vspace{-1em}
\end{table*}

\vspace{-.5em}
\subsection*{Q2: Belief-based FSCs improve inductive synthesis}
In this experiment, we run \storm for at most 1s, and use the result in \paynt.
Tab.~\ref{tab:oneway} (right) lists the results.
Our main finding is that \textcolor{blue}{\emph{inductive synthesis can find much better FSCs---and sometimes much faster---when using FSCs from belief exploration.}}
For instance, for the 4x5x2 benchmark, an FSC is obtained about six times faster while improving the value by 116\%.
On some larger models, \paynt alone struggles to find any good \fscind and using \fscbel boosts this; e.g., the value for the Refuel-20 model is raised by a factor 20 at almost no run time penalty. 
For the Tiger benchmark, a value improvement of 860\% is achieved (albeit not as good as \fscbel itself) at the expense of doubling the run time.
Thus: \emph{even a shallow exploration of the belief MDP pays off in the inductive synthesis}.
The inductive search typically profits even more when exploring the belief MDP further.
This is demonstrated, e.g., in the Rocks-12 model: using the FSC \fscbel computed using clipping (see Table~\ref{tab:oneway} (left)) enables \paynt to find FSC \fscind with the same (optimal) value 20 as \fscbel within 1s.
Similarly, for the Milos-97 model, running \storm for 45s (producing a more precise \fscbel) enables \paynt to find an FSC~\fscind achieving a better value than controllers found by \storm or \paynt alone within the timeout.
(These results are not reported in the tables.)
However, as opposed to Q1, where a better FSC \fscind naturally improves the belief MDP, longer exploring the belief MDP does not always yield a better \fscind: a larger~$\clipmdp$ with a better~\fscbel may yield a larger memory model $\mu$, thus inducing a significantly larger family where \paynt struggles to identify good FSCs.

\subsection*{Q3: The practical benefits of the symbiotic approach}
The goals of these experiments are to investigate whether the symbiotic approach improves the run time (can FSCs of a certain value be obtained faster?), the memory footprint (how is the total memory consumption affected?), the controller's value (can better FSCs be obtained with the same computational resources?) and the controller's size (are more compact FSCs obtained?). 

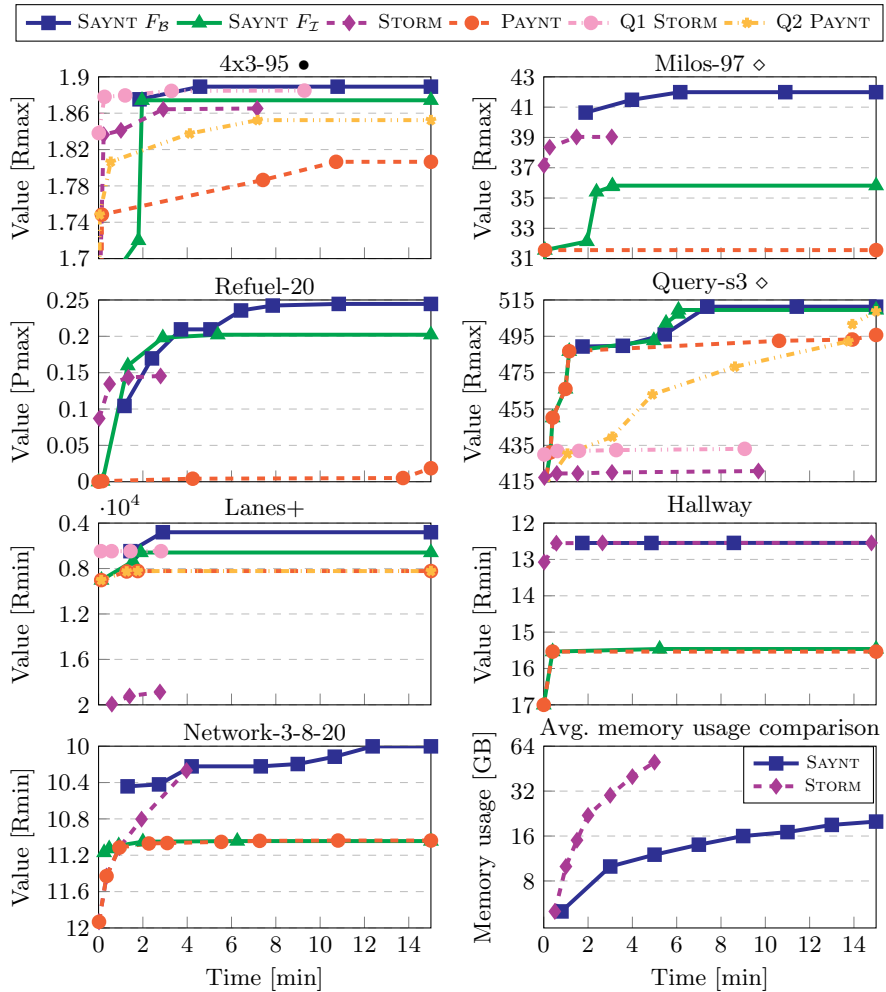
\begin{figure}[H]
\begin{tikzpicture}
\begin{groupplot}[group style={group size= 2 by 4, horizontal sep=1.5cm, vertical sep=0.55cm}, height=4cm, width=6cm]

\nextgroupplot[
    title={4x3-95 $\bullet$},
    title style={yshift=-1.8ex},
    ylabel={Value [Rmax]},
    xmin=0, xmax=15,
    ymin=1.7, ymax=1.9,
    xtick={0,2,4,6,8,10,12,14},
    xticklabels=\empty,
    ytick={1.7,1.74,1.78,1.82,1.86,1.9},
    xtick pos=left,
    ytick pos=left,
    ymajorgrids=true,
    grid style=dashed,
    y label style={at={(axis description cs:0.05,0.5)}},
    legend style ={ at={(-0.25,1.4)}, 
        anchor=north west, draw=black, nodes={scale=0.85},
        fill=white,align=center, legend columns=6,},
]

\addplot[
    color=Blue,
    every mark/.append style={solid, fill=Blue},
    mark=square*,
    line width=1.5pt,mark size=2pt,
    ]
    coordinates {
    (1.85, 1.87528)(4.58, 1.8892)(10.79, 1.8892)(15, 1.8892)
    };
    \addlegendentry{\integration \fscbel};

\addplot[
    color=Green,
    every mark/.append style={solid, fill=Green},
    mark=triangle*,
    line width=1.5pt,mark size=2pt,
    ]
    coordinates {
    (0,0)(0.05, 1.66)(1.8, 1.72)(1.96,1.8742)(15,1.8742)
    };
    \addlegendentry{\integration \fscind};

\addplot[
    color=Mulberry,
    every mark/.append style={solid, fill=Mulberry},
    mark=diamond*,
    dashed,
    line width=1.5pt,mark size=2pt,
    ]
    coordinates {
    (0.01, 1.61655)(0.23, 1.83628)(1.01, 1.84113)(2.92, 1.86433)(7.16, 1.86522)
    };
    \addlegendentry{\storm};

\addplot[
    color=RedOrange,
    every mark/.append style={solid, fill=RedOrange}, 
    mark=*,
    dashed,
    line width=1.5pt,mark size=2pt,
    ]
    coordinates {
    (0,0)(0.05, 0.66)(0.15, 1.7484)(7.42, 1.78653)(10.72, 1.80651)(15, 1.80651)
    };
    \addlegendentry{\paynt};

\addplot[
    color=Lavender,
    every mark/.append style={solid, fill=Lavender},
    mark=otimes*,
    dashdotdotted,
    line width=1.5pt,mark size=2pt,
    ]
    coordinates {
    (0.01, 1.8381)(0.27, 1.87803)(1.2, 1.87955)(3.29, 1.88455)(9.29, 1.88492)
    };
    \addlegendentry{Q1 \storm};

\addplot[
    color=Dandelion,
    every mark/.append style={solid, fill=Dandelion},
    mark=star,
    dashdotdotted,
    line width=1.5pt,mark size=2pt,
    ]
    coordinates {
    (0,0)(0.05,1.7484)(0.55,1.8065)(4.12, 1.83764)(7.18, 1.85236)(15, 1.85236)
    };
    \addlegendentry{Q2 \paynt};

\nextgroupplot[
    title={Milos-97 $\diamond$},
    title style={yshift=-1.8ex},
    ylabel={Value [Rmax]},
    xmin=0, xmax=15,
    ymin=31, ymax=43,
    xtick={0,2,4,6,8,10,12,14},
    xticklabels=\empty,
    ytick={31,33,35,37,39,41,43},
    xtick pos=left,
    ytick pos=left,
    ymajorgrids=true,
    grid style=dashed,
    y label style={at={(axis description cs:0.1,0.5)}},
]

\addplot[
    color=Blue,
    every mark/.append style={solid, fill=Blue},
    mark=square*,
    line width=1.5pt,mark size=2pt,
    ]
    coordinates {
    (1.89, 40.64122)(3.98, 41.48718)(6.16, 41.98899)(10.9, 41.98899)(15, 41.98899)
    };

\addplot[
    color=Green,
    every mark/.append style={solid, fill=Green},
    mark=triangle*,
    line width=1.5pt,mark size=2pt,
    ]
    coordinates {
    (0,0)(0.05, 31.56236)(1.95, 32.13262)(2.38, 35.4054)(3.09, 35.76289)(3.1, 35.81786)(15, 35.81786)
    };

\addplot[
    color=Mulberry,
    every mark/.append style={solid, fill=Mulberry},
    mark=diamond*,
    dashed,
    line width=1.5pt,mark size=2pt,
    ]
    coordinates {
    (0.01, 37.15222)(0.27, 38.34984)(1.48, 39.03375)(3.07, 39.03375)
    };

\addplot[
    color=RedOrange,
    every mark/.append style={solid, fill=RedOrange}, 
    mark=*,
    dashed,
    line width=1.5pt,mark size=2pt,
    ]
    coordinates {
    (0,0)(0.06, 31.56236)(15, 31.56236)
    };

\nextgroupplot[
    title={Refuel-20},
    title style={yshift=-1.8ex},
    ylabel={Value [Pmax]},
    xmin=0, xmax=15,
    ymin=0, ymax=0.25,
    xtick={0,2,4,6,8,10,12,14},
    xticklabels=\empty,
    ytick={0,0.05,0.1,0.15,0.2,0.25},
    xtick pos=left,
    ytick pos=left,
    ymajorgrids=true,
    yticklabel style={/pgf/number format/fixed},
    grid style=dashed,
    y label style={at={(axis description cs:0.05,0.5)}},
]

\addplot[
    color=Blue,
    every mark/.append style={solid, fill=Blue},
    mark=square*,
    line width=1.5pt,mark size=2pt,
    ]
    coordinates {
    (1.17, 0.10425)(2.41, 0.16963)(3.72, 0.20956)(5.05, 0.20956)(6.43, 0.23558)(7.86, 0.24232)(10.84, 0.24457)(15, 0.24457)
    };

\addplot[
    color=Green,
    every mark/.append style={solid, fill=Green},
    mark=triangle*,
    line width=1.5pt,mark size=2pt,
    ]
    coordinates {
    (0, 0)(0.18, 0.00083)(1.32, 0.15962)(2.9, 0.19814)(5.38, 0.20219)(15, 0.20219)
    };

\addplot[
    color=Mulberry,
    every mark/.append style={solid, fill=Mulberry},
    mark=diamond*,
    dashed,
    line width=1.5pt,mark size=2pt,
    ]
    coordinates {
    (0.03, 0.08705)(0.5, 0.13441)(1.35, 0.14351)(2.8, 0.14545)
    };

\addplot[
    color=RedOrange,
    every mark/.append style={solid, fill=RedOrange}, 
    mark=*,
    dashed,
    line width=1.5pt,mark size=2pt,
    ]
    coordinates {
    (0, 0)(0.17, 0.00083)(4.25, 0.00417)(13.74, 0.00512)(15, 0.01858)
    };

\nextgroupplot[
    title={Query-s3 $\diamond$},
    title style={yshift=-1.8ex},
    ylabel={Value [Rmax]},
    xmin=0, xmax=15,
    ymin=415, ymax=515,
    xtick={0,2,4,6,8,10,12,14},
    xticklabels=\empty,
    ytick={415,435,455,475,495,515},
    xtick pos=left,
    ytick pos=left,
    ymajorgrids=true,
    grid style=dashed,
    y label style={at={(axis description cs:0.07,0.5)}},
]

\addplot[
    color=Blue,
    every mark/.append style={solid, fill=Blue},
    mark=square*,
    line width=1.5pt,mark size=2pt,
    ]
    coordinates {
    (1.75, 489.38753)(3.57, 489.78038)(5.46, 496.02988)(7.38, 511.2999)(11.42, 511.31903)(15, 511.31903)
    };

\addplot[
    color=Green,
    every mark/.append style={solid, fill=Green},
    mark=triangle*,
    line width=1.5pt,mark size=2pt,
    ]
    coordinates {
    (0,0)(0.05, 413.01214)(0.3, 431.12863)(0.4, 450.22684)(0.98, 466.05536)(1.15,486.77652)(4.97, 492.68679)(5.52, 502.30175)(6.07, 507.24195)(6.08, 509.4867)(15, 509.4867)
    };

\addplot[
    color=Mulberry,
    every mark/.append style={solid, fill=Mulberry},
    mark=diamond*,
    dashed,
    line width=1.5pt,mark size=2pt,
    ]
    coordinates {
    (0.03, 417.35558)(0.59, 419.52717)(1.53, 419.5926)(3.08, 420.10552)(9.69, 420.87336)
    };

\addplot[
    color=RedOrange,
    every mark/.append style={solid, fill=RedOrange}, 
    mark=*,
    dashed,
    line width=1.5pt,mark size=2pt,
    ]
    coordinates {
    (0,0)(0.05, 413.01214)(0.3, 431.12863)(0.4, 450.22684)(0.98, 466.05536)(1.15,486.77652)(10.62, 492.51432)(13.93, 493.23566)(15, 495.74444)
    };

\addplot[
    color=Lavender,
    every mark/.append style={solid, fill=Lavender},
    mark=otimes*,
    dashdotdotted,
    line width=1.5pt,mark size=2pt,
    ]
    coordinates {
    (0.03, 429.97631)(0.61, 431.89541)(1.59, 431.95951)(3.27, 432.39931)(9.07, 433.08829)
    };

\addplot[
    color=Dandelion,
    every mark/.append style={solid, fill=Dandelion},
    mark=star,
    dashdotdotted,
    line width=1.5pt,mark size=2pt,
    ]
    coordinates {
    (0,0)(0.05, 413.01214)(1.07, 430.44502)(3.1, 439.73931)(4.91, 463.00481)(8.63,478.27243)(13.75, 492.23486)(13.93, 501.65891)(15, 508.74215)
    };

\nextgroupplot[
    title={Lanes+},
    title style={yshift=-1.8ex},
    ylabel={Value [Rmin]},
    xmin=0, xmax=15,
    ymin=4000, ymax=20000,
    xtick={0,2,4,6,8,10,12,14},
    xticklabels=\empty,
    ytick={4000,8000,12000,16000,20000},
    xtick pos=left,
    ytick pos=left,
    ymajorgrids=true,
    y dir=reverse,
    grid style=dashed,
    y label style={at={(axis description cs:0.05,0.5)}},
]

\addplot[
    color=Blue,
    every mark/.append style={solid, fill=Blue},
    mark=square*,
    line width=1.5pt,mark size=2pt,
    ]
    coordinates {
    (1.44, 6472.49336)(2.9, 4805.97703)(15, 4805.97703)
    };

\addplot[
    color=Green,
    every mark/.append style={solid, fill=Green},
    mark=triangle*,
    line width=1.5pt,mark size=2pt,
    ]
    coordinates {
    (0.13, 9019.14578)(1.55, 7359.96372)(1.91, 6591.11316)(15, 6591.11316)
    };

\addplot[
    color=Mulberry,
    every mark/.append style={solid, fill=Mulberry},
    mark=diamond*,
    dashed,
    line width=1.5pt,mark size=2pt,
    ]
    coordinates {
    (0.11, 21286.62547)(0.6, 19936.40517)(1.4, 19230.93944)(2.77, 18870.51119)
    };

\addplot[
    color=RedOrange,
    every mark/.append style={solid, fill=RedOrange}, 
    mark=*,
    dashed,
    line width=1.5pt,mark size=2pt,
    ]
    coordinates {
    (0.12, 9019.14578)(1.27, 8260.19657)(1.77, 8223.35307)(15, 8223.35307)
    };

\addplot[
    color=Lavender,
    every mark/.append style={solid, fill=Lavender},
    mark=otimes*,
    dashdotdotted,
    line width=1.5pt,mark size=2pt,
    ]
    coordinates {
    (0.11, 6472.62324)(0.59, 6472.49336)(1.44, 6472.18581)(2.82, 6471.86978)
    };

\addplot[
    color=Dandelion,
    every mark/.append style={solid, fill=Dandelion},
    mark=star,
    dashdotdotted,
    line width=1.5pt,mark size=2pt,
    ]
    coordinates {
    (0.12, 9019.14578)(1.27, 8260.19657)(1.77, 8223.35307)(15, 8223.35307)
    };

\nextgroupplot[
    title={Hallway},
    title style={yshift=-1.8ex},
    ylabel={Value [Rmin]},
    xmin=0, xmax=15,
    ymin=12, ymax=17,
    xtick={0,2,4,6,8,10,12,14},
    xticklabels=\empty,
    ytick={12,13,14,15,16,17},
    xtick pos=left,
    ytick pos=left,
    ymajorgrids=true,
    y dir=reverse,
    grid style=dashed,
    y label style={at={(axis description cs:0.1,0.5)}},
]

\addplot[
    color=Blue,
    every mark/.append style={solid, fill=Blue},
    mark=square*,
    line width=1.5pt,mark size=2pt,
    ]
    coordinates {
    (1.74, 12.54746)(4.86, 12.54746)(8.58, 12.54528)(17.73, 12.54527)
    };

\addplot[
    color=Green,
    every mark/.append style={solid, fill=Green},
    mark=triangle*,
    line width=1.5pt,mark size=2pt,
    ]
    coordinates {
    (0,17)(0.4, 15.53578)(5.23, 15.46362)(15, 15.46362)
    };

\addplot[
    color=Mulberry,
    every mark/.append style={solid, fill=Mulberry},
    mark=diamond*,
    dashed,
    line width=1.5pt,mark size=2pt,
    ]
    coordinates {
    (0.02, 13.07497)(0.57, 12.55379)(2.65, 12.54531)(14.79, 12.54531)
    };

\addplot[
    color=RedOrange,
    every mark/.append style={solid, fill=RedOrange}, 
    mark=*,
    dashed,
    line width=1.5pt,mark size=2pt,
    ]
    coordinates {
    (0,17)(0.4, 15.53578)(15, 15.53578)
    };

\nextgroupplot[
    title={Network-3-8-20},
    title style={yshift=-1.8ex},
    xlabel={Time [min]},
    ylabel={Value [Rmin]},
    xmin=0, xmax=15,
    ymin=10, ymax=12,
    xtick={0,2,4,6,8,10,12,14},
    ytick={10,10.4,10.8,11.2,11.6,12},
    xtick pos=left,
    ytick pos=left,
    ymajorgrids=true,
    y dir=reverse,
    grid style=dashed,
    x label style={at={(axis description cs:0.5,0.05)}},
    y label style={at={(axis description cs:0.05,0.5)}},
    legend style ={ at={(1.4,0.75)}, 
        anchor=north west, draw=black, 
        fill=white,align=left},
]

\addplot[
    color=Blue,
    every mark/.append style={solid, fill=Blue},
    mark=square*,
    line width=1.5pt,mark size=2pt,
    ]
    coordinates {
    (1.31, 10.44199)(2.73, 10.42043)(4.2, 10.2219)(7.32, 10.2219)(8.99, 10.19629)(10.66, 10.11574)(12.37, 10.0018)(15, 10.0018)
    };

\addplot[
    color=Green,
    every mark/.append style={solid, fill=Green},
    mark=triangle*,
    line width=1.5pt,mark size=2pt,
    ]
    coordinates {
    (0.25, 11.17)(0.48, 11.12752)(0.91, 11.09404)(2.0, 11.05084)(6.26, 11.04406)(15,11.04406)
    };

\addplot[
    color=Mulberry,
    every mark/.append style={solid, fill=Mulberry},
    mark=diamond*,
    dashed,
    line width=1.5pt,mark size=2pt,
    ]
    coordinates {
    (0.02, 11.93271)(0.36, 11.429)(0.94, 11.11544)(1.94, 10.80298)(3.97, 10.26813)
    };

\addplot[
    color=RedOrange,
    every mark/.append style={solid, fill=RedOrange}, 
    mark=*,
    dashed,
    line width=1.5pt,mark size=2pt,
    ]
    coordinates {
    (0.02, 11.93271)(0.36, 11.429)(0.94, 11.11544)(2.27, 11.07074)(3.1, 11.0659)(5.54, 11.05325)(7.27, 11.04164)(10.81, 11.03808)(15, 11.03808)
    };

\nextgroupplot[
    title={Avg.~memory usage comparison},
    title style={yshift=-1.8ex},
    xlabel={Time [min]},
    ylabel={Memory usage [GB]},
    xmin=0, xmax=15,
    ymin=0, ymax=64,
    xtick={0,2,4,6,8,10,12,14},
    ytick={0,8,16,32,64},
    xtick pos=left,
    ytick pos=left,
    ymode=log,
    log ticks with fixed point,
    log basis y=2,
    ymajorgrids=true,
    grid style=dashed,
    x label style={at={(axis description cs:0.5,0.05)}},
    y label style={at={(axis description cs:0.1,0.5)}},
    legend style ={ at={(0.6,1),}, 
        anchor=north west, align=center, nodes={scale=0.75, transform shape}},
]

\addplot[
    color=Blue,
    every mark/.append style={solid, fill=Blue},
    mark=square*,
    line width=1.5pt,mark size=2pt,
    ]
    coordinates {
    (0,0)(0.8,5)(3, 10)(5, 12)(7,14)(9,16)(11,17)(13,19)(15,20)
    };
    \addlegendentry{\integration}

\addplot[
    color=Mulberry,
    every mark/.append style={solid, fill=Mulberry},
    mark=diamond*,
    dashed,
    line width=1.5pt,mark size=2pt,
    ]
    coordinates {
    (0,0)(0.5,5)(1,10)(1.5,15)(2,22)(3,30)(4,40)(5,50)
    };
    \addlegendentry{\storm}

\end{groupplot}
\end{tikzpicture}
\vspace{-0.8em}
\caption{Value of the generated FSCs over time. The last graph shows the average memory usage of \storm and \integration. 
The lines ending before the timeout indicate 
that the 64GB memory limit was hit.
$\bullet$ indicates that \paynt and \integration
synthesised posterior-aware FSCs. $\diamond$ indicates that \integration ran with $t_{\inductive}=$90s. 
}
\label{fig:integration-graphs}
\end{figure}

\paragraph{Value of the synthesised FSCs.}
Figure~\ref{fig:integration-graphs} plots the value of the FSCs produced by \storm, \paynt, and \integration versus the computation time. 
Note that for maximal objectives, the aim is to obtain a high value (the first 4 plots) whereas for minimal objectives a lower value prevails.
From the plots, it follows that \textcolor{blue}{\emph{the FSCs from the symbiotic approach are superior in value to the ones obtained by the standalone approaches.}}
The relative improvement of the value of the resulting FSCs differs across individual models, similar to the trends in Q1 and Q2.
When comparing the best FSC found by \storm or \paynt alone with the best FSC found by \integration, the improvement ranges from negligible (4x3-95) to around 3\%-7\% (Netw-3-8-20, Milos-97, Query-s3) and sometimes goes over 40\% (Refuel-20, Lines+). We note that the distance to the (unknown) optimal values remains unclear.
The FSC value never decreases but sometimes does also not increase, as indicated by 
Hallway and Rocks-12 (see also Q2). 
Our experiments (see Tab.~\ref{tab:integration}) also indicate that the improvement over the baseline algorithms is typically  more significant in the larger variants of the models.
Furthermore, the plots in Fig.~\ref{fig:integration-graphs} also include the FSC value by the one-shot combination of \storm and \paynt.
We see that \textcolor{blue}{\emph{\integration can improve the FSC value over the one-shot combination.}}
This is illustrated in, e.g., the 4x3-95 and Lanes+ benchmarks, see the 1st and 3rd plots in Fig.~\ref{fig:integration-graphs} (left).

\paragraph{Total synthesis time.}
\integration initially needs some time for the first iteration (one inductive and one belief phase) in Alg.~\ref{alg:integrated} and thus during the beginning of the synthesis process, the standalone tools may 
provide FSCs of a certain value faster. \textcolor{blue}{\emph{After the first iteration, however, \integration typically provides better FSCs in a shorter time.}}
For instance, for the Refuel-20 benchmark \integration swiftly overtakes \storm after the first iteration. 
The only exception is Rocks-12 (discussed before), where \integration with the default settings needs significantly more time than \storm to obtain an FSC of the same value.

\paragraph{Memory footprint.}
Belief exploration typically has a large memory footprint: \storm quickly hits the 64GB memory limit on exploring the belief MDP.
\textcolor{blue}{\emph{\integration reduces the memory footprint of \storm alone by a factor 3 to 4}}, see the bottom right plot of Fig.~\ref{fig:integration-graphs}. The average memory footprint of running \paynt standalone quickly stabilises around 700MB.
The memory footprint of \integration is thus dominated by the restricted exploration of the belief MDP.

\begin{table}[t]
 \setlength{\tabcolsep}{1.2pt}
    \centering
    \begin{tabular}{|c||c|c|c|c|c|c|c|c|}
    \hline
       Models:    & Lanes+  & Hallway & Netw-3-8-20& Query-s3$\diamond$& Refuel-06 & Drone-8-2 & Refuel-20    \\ \hline
      \fscbel   & 4805/8.1k  & 12.55/2k & 10/40k& 511.32/7.7k& 0.67/84 &0.96/237k & 0.24/1.5k  \\
      \fscind   & 6591/34  & 15.46/86 & 11.04/4.8k& 509.49/26& 0.67/156 &0.90/6.4k & 0.2/362 \\ \hline
      
    \end{tabular}
    \vspace{0.5em}
    \caption{Trade-offs between the value and size in the resulting FSCs \fscind and \fscbel found by \integration. Each cell reports value/size. The first three models have a minimising objective.
    $\diamond$ indicates that \integration ran with $t_{\inductive}=$90s.
    }
    \label{tab:size}
    \vspace{-2em}
\end{table}

\paragraph{The size of the synthesised FSCs.}
For selected models, Tab.~\ref{tab:size} shows the trade-offs between the value and size of the resulting FSCs \fscind and \fscbel found by \integration.
The experiments show that \textcolor{blue}{\emph{the FSCs \fscind provided by inductive synthesis are typically about one to two orders of magnitude smaller than the belief-based FSCs \fscbel with only a small penalty in their values}}. There are models (e.g. Refuel-06) where a very small \fscbel, having even slightly smaller size than \fscind, does exist. 
The integration mostly reduces the size of \fscbel due to the better approximation of the belief MDP by up to a factor of two.
This reduction has a negligible effect on the size of \fscind.
This observation further strengthens the usefulness of \integration that jointly improves the value of \fscind and \fscbel. 
Hence, \integration gives users a unique opportunity to run a single, time-efficient synthesis and select the FSC according to the trade-off between its value and size.

\paragraph{Customising the \integration setup.} 
In contrast to the standalone approaches as well as to the one-way integrations presented in Q1 and Q2, \textcolor{blue}{\emph{\integration provides a single synthesis method that is efficient for a general class of models without tuning its parameters}}.
Naturally, adjusting the parameters to individual benchmarks can further improve the quality of the computed controllers: captions of Fig.~\ref{fig:integration-graphs} and Tab.~\ref{tab:size} describe which non-default settings were used for selected models.

\subsection*{Additional results}

In Tab.~\ref{tab:integration}, we compare values and sizes of FSCs synthesised by the particular methods on a broader range of benchmarks.
We can see that FSCs \fscind obtained by \integration achieve better values than the controllers computed by \paynt; size-wise, these better FSCs of \integration are similar or only slightly bigger.
Meanwhile, for FSCs \fscbel obtained by \integration, we sometimes observe a significant size reduction while still improving the value compared to the FSCs produced by \storm. 
Two models are notable: On Drone-8-2, \integration obtains 50\% smaller \fscbel while having a 41\% better value. On Network-3-8-20, the size of \fscbel is reduced by 40\% while again providing better value.

In the following, we further discuss the impact of non-default settings for selected benchmarks, as presented in Tab.~\ref{tab:integration}.
For instance, using posterior-aware FSCs generally significantly slows down the synthesis process, however, for Network and 4x3-95, it helps improve the value of the default posterior-unaware FSCs by 2\% and 4\%, respectively. 
For the former model, a better \fscind also improves \fscbel by about a similar value. 
In some cases, e.g. for Query-s3, it is beneficial to increase the parameter $t_{\inductive}$, giving \paynt enough time to search for a good FSC \fscind (the relative improvement is 6\%), which also improves the value of the resulting FSC \fscbel by about a similar value.
Tuning $t_{\inductive}$ and $t_{\belief}$ can also have an impact on the value-size trade-off, as seen in the Milos-97 model, where setting longer timeout $t_{\inductive}$ results in finding a 2\% better \fscbel with 130\% size increase.
A detailed analysis of the experimental results suggests that usually, it is more beneficial to invest time into searching for good \fscind that is used to compute better cut-off values, rather than into deeper exploration of belief MDP.
However, the timeouts still need to allow for multiple subsequent iterations of the algorithm in order to utilise the full potential of the~symbiosis.

\section{Conclusion and Future Work}
We proposed \integration, a symbiotic integration of the two main approaches for controller synthesis in POMDPs. Using a wide class of models, we demonstrated that \integration substantially improves the value of the resulting controllers and provides an any-time, push-button synthesis algorithm allowing users to select the controller based on the trade-off between its value and size, and the synthesis~time. 

In future work, we plan to explore if the inductive policy synthesis can also be successfully combined with point-based approximation methods, such as SARSOP, and on discounted reward properties. A preliminary comparison on discounting properties 
provides two interesting observations: 1) For models with large reachable belief space and discount factors (very) close to one, SARSOP typically fails to update its initial \emph{alpha-vectors} and thus produces low-quality controllers. In these cases, SAYNT outperforms SARSOP. 2) For common discount factors, SARSOP beats SAYNT on the majority of benchmarks. This is not surprising, as the MDP engine underlying SAYNT does not natively support discounting and instead computes a much harder fixed point. See~\cite{hartmanns2023practitioner}, for a recent discussion on the differences between discounting and not discounting.

\vfill

\begin{table}[t]
\renewcommand{\arraystretch}{0.95}
\makebox[\textwidth][c]{
\scalebox{1}{
\begin{tabular}{|cc||rr||r|r||r|r||r|r|r|r|}
\hline

\multicolumn{2}{|c||}{Benchmark} & 
\multicolumn{2}{c||}{Model Size} & 
\multicolumn{2}{c||}{\paynt} & 
\multicolumn{2}{c||}{\storm} &  
\multicolumn{4}{c|}{\integration} 
\\

\multicolumn{1}{|c}{Model} & \multicolumn{1}{c||}{Spec.} & 
\multicolumn{1}{c}{$S$/$\Sigma \Act$} & \multicolumn{1}{c||}{$Z$} & 
\multicolumn{1}{c|}{\fscind}& \multicolumn{1}{c||}{Size} &
\multicolumn{1}{c|}{\fscbel}& \multicolumn{1}{c||}{Size} &
\multicolumn{1}{c|}{\fscbel} & \multicolumn{1}{c|}{Size} &
\multicolumn{1}{c|}{\fscind} & \multicolumn{1}{c|}{Size} 
\\ \hline \hline

 & 
\multirow{4}{*}{$R_\mathrm{max}$} & 
 & 
\multirow{4}{*}{$9$} & 
 & \multirow{4}{*}{36} &
 & \multirow{4}{*}{999}  &
$\mathbf{1.89}\bullet$ & 968 &
$1.87\bullet$ & 126 
\\ 

4x3& 
& 
$22$ &
&  
$1.81$ &  &
$1.87$ &  &
$283s$ & &
$120s$ & 
\\ \cline{9-12}

95& 
 & 
$82$ & 
 & 
$764s$ & &
$414s$ &  &
$\mathbf{1.89}$ & 869 &
$1.79$ & 36 
\\ 

& 
& 
 &
&  
 &  &
 & &
$303s$ & &
$678s$ & 
\\ \hline

4x5x2 & 
\multirow{2}{*}{$R_\mathrm{max}$} & 
$79$ & 
\multirow{2}{*}{$7$} & 
$0.94$ & 26 &
$2.08$ & 102 &
$2.08$ & 102 &
$2.03$ & 38 
\\ 

95& 
& 
$310$ &
&  
$305s$ &  &
$3s$ & &
$71s$ & &
$378s$ & 
\\ \hline

 & 
\multirow{4}{*}{$P_\mathrm{max}$} & 
 & 
\multirow{4}{*}{$384$} & 
 & \multirow{4}{*}{768}  &
 & \multirow{4}{*}{170k} &
$\mathbf{0.89}\bullet$ & 169k &
$0.87\bullet$ & 2.5k 
\\ 

Drone & 
& 
$1226$ &
&  
$0.87$ &  &
$0.84$ & &
$390s$ & &
$453s$ & 
\\ \cline{9-12}

4-1 & 
& 
$3026$ & 
 & 
$665s$ &  &
$110s$ &  &
$\mathbf{0.89}$ & 176k &
$0.79$ & 922 
\\ 

 & 
& 
 &
&  
 &  &
 & &
$180s$ & &
$45s$ & 
\\ \hline

Drone & 
\multirow{2}{*}{$P_\mathrm{max}$} & 
$1226$ & 
\multirow{2}{*}{$761$} & 
$0.95$ & 1.5k  &
$0.95$ & 135k &
$\mathbf{0.97}$ & 140k &
$0.94$ & 1.5k  
\\ 

4-2 & 
& 
$3026$ &
&  
$900s$ &  &
$110s$ & &
$194s$ & &
$1s$ & 
\\ \hline

Drone & 
\multirow{2}{*}{$P_\mathrm{max}$} & 
$13$k & 
\multirow{2}{*}{$3195$} & 
$0.9$ & 6.4k  &
$0.68$ & 280k &
$\mathbf{0.96}$ & 140k &
$0.9$ & 6.4k 
\\ 

8-2 & 
& 
$32$k &
&  
$260s$ &  &
$98s$ & &
$247s$ & &
$30s$ & 
\\ \hline

\multirow{2}{*}{Hallway} & 
\multirow{2}{*}{$R_\mathrm{min}$} & 
$61$ & 
\multirow{2}{*}{$23$} & 
$15.54$ & 66  &
$12.55$ & 1.9k  &
$12.55$ & 1.8k &
$15.46$ & 86  
\\ 

& 
& 
$301$ &
&  
$26s$ &  &
$916s$ &  &
$263s$ & &
$293s$ & 
\\ \hline

\multirow{2}{*}{Lanes+}& 
\multirow{2}{*}{$R_\mathrm{min}$} & 
$2741$ & 
\multirow{2}{*}{$11$} & 
$8223$ & 42  &
$18870$ & 8.1k  &
$\mathbf{4805}$ & 8.1k &
$6591$ & 34 
\\ 

& 
& 
$5289$ &
&  
$118s$ &  &
$376s$ & &
$173s$ &  &
$114s$ &  
\\ \hline

\multirow{4}{*}{Milos-97} & 
\multirow{4}{*}{$R_\mathrm{max}$} & 
 & 
\multirow{4}{*}{$11$} & 
 & \multirow{4}{*}{40}  &
 & \multirow{4}{*}{823}  &
$\mathbf{41.99}\diamond$ & 692 &
$35.82\diamond$ & 40 
\\ 

& 
& 
$165$ &
&  
$31.56$ &  &
$39.03$ &  &
$370s$ &  &
$185s$ & 
\\ \cline{9-12}

& 
 & 
$980$ & 
 & 
$4s$ &  &
$88s$ &   &
$\mathbf{41.55}$ & 290 &
$35.41$ & 40  
\\ 

& 
& 
 &
&  
 &  &
 &  &
$270s$ &  &
$114s$ & 
\\ \hline

\multirow{4}{*}{Network} & 
\multirow{4}{*}{$R_\mathrm{max}$} & 
 & 
\multirow{4}{*}{$5$} & 
 & \multirow{4}{*}{22}  &
 & \multirow{4}{*}{2.4k}  &
$\mathbf{289.18}\bullet$ & 2k &
$287.23\bullet$ & 54 
\\ 

& 
& 
$19$ &
&  
$280.33$ &  &
$209.71$ & &
$395s$ & &
$106s$ & 
\\ \cline{9-12}

 & 
 & 
$70$ & 
 & 
$38s$ &   &
$110s$ &  &
$\mathbf{284.51}$ & 1.8k  &
$280.33$ & 22 
\\ 

& 
& 
&
&  
 &  &
 & &
$85s$ & &
$41s$ & 
\\ \hline

Netw & 
\multirow{2}{*}{$R_\mathrm{min}$} & 
$4589$ & 
\multirow{2}{*}{$1173$} & 
$4.24$ & 2.3k  &
$3.21$ & 34k  &
$\mathbf{3.2}$ & 23k &
$4.19$ & 2.5k 
\\ 

2-8-20 & 
& 
$6973$ &
&  
$914s$ &  &
$11s$ & &
$71s$ & &
$211s$ & 
\\ \hline

Netw & 
\multirow{2}{*}{$R_\mathrm{min}$} & 
$17$k & 
\multirow{2}{*}{$2205$} & 
$11.04$ & 4.4k  &
$10.27$ & 64k  &
$\mathbf{10}$ & 38k &
$11.04$ & 4.8k 
\\ 

3-8-20 & 
& 
$30$k &
&  
$638s$ &  &
$238s$ &  &
$742s$ & &
$379s$ & 
\\ \hline

 & 
\multirow{4}{*}{$R_\mathrm{max}$} & 
 & 
\multirow{4}{*}{$6$} & 
 & \multirow{4}{*}{28}  &
 &  \multirow{4}{*}{12.9k} &
$\mathbf{511.32}\diamond$ & 7.7k &
$509.49\diamond$ & 26 
\\ 

Query & 
& 
$108$ &
&  
$502.3$ &  &
$420.11$ & &
$566s$ & &
$362s$ & 
\\ \cline{9-12}

s3 & 
 & 
$320$ & 
 & 
$931s$ &  &
$184s$ &   &
$482.21$ & 7.7k &
$478.59$ & 28 
\\ 

 & 
& 
 &
&  
 &  &
& &
$700s$ & &
$610s$ & 
\\ \hline

Refuel & 
\multirow{2}{*}{$P_\mathrm{max}$} & 
$208$ & 
\multirow{2}{*}{$50$} & 
$0.35$ & 100  &
$0.67$ & 343 &
$0.67$ & 84 &
$0.67$ & 156 
\\ 

06 & 
& 
$565$ &
&  
${<}1s$ &  &
$182s$ &  &
$178s$ & &
$84s$ & 
\\ \hline

Refuel & 
\multirow{2}{*}{$P_\mathrm{max}$} & 
$470$ & 
\multirow{2}{*}{$66$} & 
$0.32$ & 132  &
$0.44$ &  534 &
$\mathbf{0.45}$ & 140 &
$0.3$ & 142 
\\ 

08 & 
& 
$1431$ &
&  
$253s$ &  &
$96s$ & &
$186s$ & &
$84s$ & 
\\ \hline

Refuel & 
\multirow{2}{*}{$P_\mathrm{max}$} & 
$6834$ & 
\multirow{2}{*}{$174$} & 
$0.02$ & 348  &
$0.15$ & 1.2k &
$\mathbf{0.24}$ & 1.5k &
$0.2$ & 360 
\\ 

20 & 
& 
$24$k &
&  
$922s$ &  &
$468s$ &  &
$386s$ & &
$173s$ & 
\\ \hline

Rocks & 
\multirow{2}{*}{$R_\mathrm{min}$} & 
$6553$ & 
\multirow{2}{*}{$1645$} & 
$42$ & 3.3k &
$\mathbf{20}*$ & 115 &
$20*$ & 115 &
$20*$ & 3.3k 
\\ 

12 & 
& 
$32$k &
&  
${<}1s$ &  &
$15s$ &  &
$235s$ &  &
$236s$ & 
\\ \hline

Tiger & 
\multirow{2}{*}{$R_\mathrm{max}$} & 
$14$ & 
\multirow{2}{*}{$7$} & 
$7.93$ & 34  &
$50.38$ & 58 &
$50.38$ & 58 &
$31.61$ & 48 
\\ 

95 & 
& 
$50$ &
&  
$547s$ &  &
${<}1s$ & &
$71s$ & &
$513s$ & 
\\ \hline

\end{tabular}
}
}
\vspace{0.5em}
\caption{The quality and size of resulting FSCs provided by \paynt, \storm, and \integration within the \mbox{15-minute} timeout. The run times indicate the time needed to find the best FSC. Non-default settings: $*$ marks experiments where clipping was enabled, $\bullet$ marks experiments where \tool{PAYNT} synthesised posterior-aware FSCs, $\diamond$ marks experiments where integration parameter $t_{\inductive}$ was set to 90 seconds.}
\label{tab:integration}
\end{table}

\clearpage
\bibliographystyle{splncs04}
\bibliography{main}

\clearpage
\appendix

\section{MDP abstraction for general FSCs}
\label{app:preciseabstraction}

\paragraph{Design space.}
It is convenient to have a concise representation of a family of FSCs.
Assume POMDP $\pomdp$, memory model $\mu$ and a family $\ffscmu$ of $\mu$-FSCs with the set $N = \{n_0,\dots,n_{k-1}\}$ of memory nodes, where $k=\max_{z \in Z} \mu(z)$.
Let $\nodes{z} \coloneqq \{n_0,\dots,n_{\mu(z)-1}\}$ denote the set of memory nodes available to observation $z$.
Family $\ffscmu$ is parameterised by the choice of the action function $\gamma \colon N \times Z \rightarrow \Act$ and the update function $\delta \colon N \times Z \times Z \rightarrow N$.
Thus, to represent $\ffscmu$, for every $z,z' \in Z$ and for each $n \in \nodes{z}$, we introduce parameters $\gamma_{n}^z \in \Act$, $\delta_{n}^{z,z'} \in \nodes{z'}$.
The sets
\[
\Gamma_\mu \coloneqq \!\! \prod_{z \in Z} \prod_{n \in \nodes{z}}\!\!  \left\{ \gamma_n^z=\act \mid \act \in \Act \right\}
\quad \text{ and }\]

\[\Delta_\mu \coloneqq \!\! \prod_{z \in Z} \prod_{n \in \nodes{z}} \prod_{z' \in Z}\!\! \left\{ \delta_{n}^{z,z'}=n' \mid n' \in \nodes{z'} \right\}\]
collect possible assignments of parameters related to the functions $\gamma$ and $\delta$, respectively. Thus, the set of policies is represented by the \emph{design space} $\designmu \coloneqq \Gamma_\mu \times \Delta_\mu$.
The goal of the inductive synthesis is now to select the assignment of all parameters $\gamma_n^z$ and $\delta_n^{z,z'}$ so that the corresponding FSC $\fscT$ induces the optimal Markov chain.

\paragraph{Refined MDP abstraction.}
Let $post(s,\act) \coloneqq \{O(s') \mid s' \in \supp(\mpm(s,\act)) \}$ denote the set of posterior observations available when executing $\act \in \Act$ in $s \in S$. 
Then, if $\act$ is executed in state $s$ and $n \in \nodes{O(s)}$ is the current memory node, then the following parameter assignments
\[
\Delta_\mu[s,n,\act] \coloneqq \prod_{z' \in post(s,\act)} \left\{ \delta_n^{O(s),z'}=n' \mid n' \in \nodes{z'} \right\}
\]
are relevant to determine the new memory node.
The following definition refines the earlier naive version in Def.~\ref{def:mdpapproxnaive}.

\begin{definition}
\emph{MDP abstraction} for POMDP $\pomdp$ and a family $\ffscmu$ of $\mu$-FSCs is an MDP $\textsf{MDP}\left( \ffscmu \right) \coloneqq \left(S\times N, (\sinit,\ninit), \Act \times \Delta_\mu, \mpm^\ffscmu \right)$
with the transition function
$$\mpm^\ffscmu ( (s,n), (\act,\delta) ) \coloneqq \left\{(s',n') \mapsto \iversion{ \delta_n^{O(s),O(s')}=n' } \cdot \mpm(s,\act,s') \right\}$$
\end{definition}
Notice that in this MDP abstraction state $(s,n)$ has only $\sum_{\act \in Act}\left| \Delta_\mu[s,n,\act] \right|$ unique actions.


\section{Sizes of FSC of various sub-classes}
\label{app:sizes}

Assume a POMDP $\pomdp$, a $k$-FSC $F = \fscT$ with the set $N = \{n_0,\dots,n_{k-1}\}$ of memory nodes and an induced MC $\pomdp^F = (S \times N, (\sinit,\ninit), \{\act\}, \mpm^F)$.
Recall that in order to avoid an explicit encoding of $\delta$ requiring ${k \cdot |Z|^2}$ space, we use sets
$post(n,z) \coloneqq \left\{O(s') \mid \exists s \in S_z \colon (s',\cdot) \in \supp(\mpm^F((s,n),\act)) \right\}$
of posterior observations available when making a transition in states $(s,n)$ of $\pomdp^F$ with prior observation $z$. Then, $\delta(n,z,\cdot)$ can be encoded as a list $\left\{(z',\delta(n,z,z') \mid z' \in post(n,z) \right\}$ of posterior-node pairs. Thus,
$$\size(\delta) = \sum_{n \in N} \sum_{z \in Z} 2 \cdot |post(n,z)| = 2 \cdot \sum_{n \in N} \sum_{z \in Z} |post(n,z)|.$$

Using memory model $\mu$ refines the generic $k$-factor: for observation $z \in Z$ there are now not $k$ but only $\mu(z)$ distinct memory nodes. Therefore,
\begin{align*}
\size(\gamma) &= \sum_{z \in Z} \sum_{i=0}^{\mu(z)-1} 1 = \sum_{z \in Z}\mu(z) \\
\size(\delta) &= 2 \cdot \sum_{z \in Z} \sum_{i=0}^{\mu(z)-1} |post(n_i,z)|
\end{align*}
If $\mu$-FSC is posterior-unaware, i.e.~$\delta(n_i,z,z')$ is the same for all $z' \in Z$, then for each $z \in Z$ and $n_i \in \left\{n_0,\dots,n_{\mu(z)-1} \right\}$ it is sufficient to store a single value $\delta(n_i,z,\cdot)$. Thus, 
$$\size(\delta) = \sum_{z \in Z} \sum_{i=0}^{\mu(z)-1} 1 = \sum_{z \in Z}\mu(z)$$
and $\size(\gamma) = \sum_{z \in Z}\mu(z)$, as before.

\vspace{0.5em}

Finally, assume a composite FSC $\fscbel = (\exploredbel \cup N, \binit, \gamma, \delta)$ obtained after applying $\fscind = (N,n_0,\gamma_{\inductive}, \delta_{\inductive})$ at frontier states. Recall that each non-frontier state $b \in \exploredbel$ is associated with the unique prior observation $O(b)$.
Therefore, for every $b \in \exploredbel$ we must store exactly 1 action and a list $\left\{(z',\delta(b,O(b),z') \mid z' \in post(b,O(b)) \right\}$ of posterior-belief  pairs. Finally, we must also account for the size of FSC $\fscind$ used for cut-offs. Overall, we obtain:
\begin{align*}
\size(\gamma) &= \size(\gamma_{\inductive}) + \sum_{b \in \exploredbel} 1 = \size(\gamma_{\inductive}) + |\exploredbel| \\
\size(\delta) &= \size(\delta_{\inductive}) + \sum_{b \in \exploredbel} 2 \cdot \left|post(b,O(b))\right| = \size(\delta_{\inductive}) + 2 \cdot \sum_{b \in \exploredbel} \left|post(b,O(b))\right|
\end{align*}

\vfill

\section{New POMDP Lanes+}
\label{app:lanes}

\begin{figure}[!t]
\centering
\begin{subfigure}{.35\textwidth}
\centering
\includegraphics[width=0.65\linewidth]{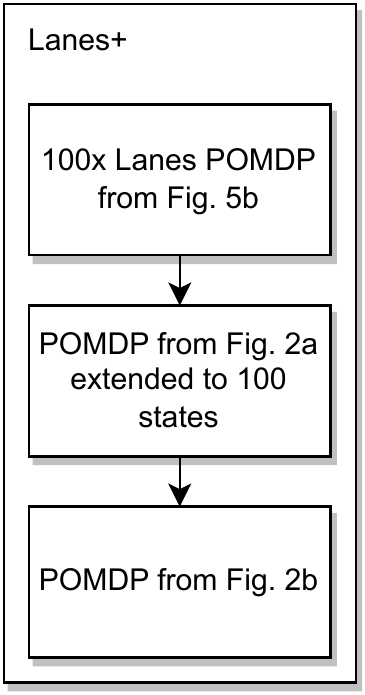}
\caption{}
\label{fig:lanes-plus}
\end{subfigure}%
\begin{subfigure}{.65\textwidth}
\centering
\includegraphics[width=0.9\linewidth]{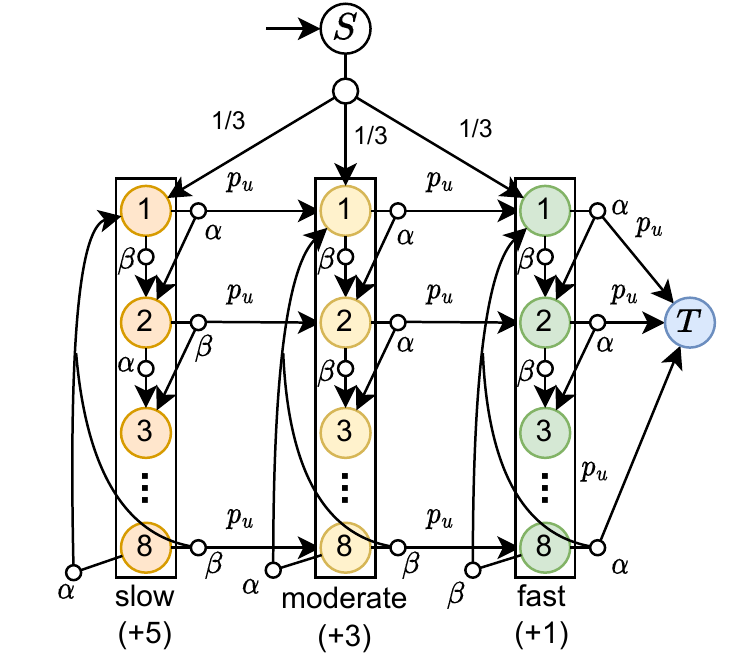}
\caption{}
\label{fig:lanes}
\end{subfigure}%
\caption{(a) Overall structure of the Lanes+ POMDP. (b) The Lanes POMDP. When performing the upgrading action (e.g.~action $\alpha$ in the first state of the slow lane), the lane is upgraded with probability $p_u$; with probability $1{-}p_u$ the agent moves to the next state of the lane.}
\label{fig:lanes-overview}
\end{figure}

This section describes the new model, Lanes+, used in our experimental evaluation. 
Fig.~\ref{fig:lanes-plus} illustrates the structure of the Lanes+ model: it is a sequential composition of a Lanes POMDP (described below) repeated 100 times, followed by the POMDP from Fig.~\ref{fig:storm-problem} extended to 100 states, followed by the POMDP from Fig.~\ref{fig:paynt-problem}.
The core component~--~Lanes model~--~was designed with two main goals in mind: i) the optimal FSC \fscind requires several memory nodes and ii) the model can be easily scaled up such that an exhaustive policy search is not feasible.
When combining Lanes with POMDPs from Fig~\ref{fig:problems}, we obtain a model the analysis of which is unfeasible for standalone methods and requires their two-way integration.

Fig.~\ref{fig:lanes} depicts the structure of the Lanes model.
Similarly to POMDPs presented in Fig.~\ref{fig:problems}, assume an agent that attempts to reach the target state~$T$ as fast as possible.
In order to reach $T$, the agent must cross three \emph{lanes}: slow, moderate and fast.
The agent initially starts in a random lane chosen uniformly (i.e.~with probability 1/3).
Each lane consists of 8 states and in each state, the agent has two actions available: $\alpha$ and $\beta$. One of these actions is an upgrading action which moves the agent with probability $p_u$ to the next lane (or to the target state when performed in the fast lane); with probability $1{-}p_u$ the agent moves to the next state of the same lane. The other action is a stalling action: it moves the agent to the next state of the same lane with probability 1.
Which of the actions $\alpha,\beta$ is the upgrading one differs depending on the lane and the current position in the lane.
Additionally, the lanes are circular: when residing in the last state of the lane, failing to upgrade the lane or performing the stalling action moves the agent back to the beginning of the current lane.
The only available observation for the agent is its current lane.

There are two differences between individual lanes. First, their speed: performing any action in the slow lane takes 5 time units, 3 for the moderate lane and 1 for the fast lane.
Second, the distribution of upgrading/stalling actions is also different in each lane.
For instance, in the slow lane, every other state has action $\alpha$ as the upgrading one; in the moderate one, states $1,2$ and $5,6$ have $\alpha$ as the upgrading action, etc.
Due to this irregularity of upgrading actions as well as due to the fact that only lanes are observable, but not individual states, the best policy must keep track of the current position in the lane, implying that 8 memory nodes are needed to encode the optimal FSC.
The corresponding family $\ffscmu$ contains $\approx\!10^{43}$ controllers, which is unattainable for inductive search.
Meanwhile, the belief-based analysis is straightforward since the belief MDP $\bmdp$ is finite.

\end{document}